\documentclass[11pt,reqno]{article}
\usepackage[T1]{fontenc}
\usepackage[latin9]{inputenc}
\usepackage[a4paper]{geometry}
\usepackage{float}
\usepackage{amsmath}
\usepackage{amsthm}
\usepackage{amssymb}
\usepackage{graphicx}
\usepackage{setspace}
\makeatletter

\providecommand{\tabularnewline}{\\}

\usepackage{amsfonts}

\newtheorem{thm}{Theorem}
\newtheorem{lem}{Lemma}[section]
\newtheorem{prop}{Proposition}
\newtheorem{asm}{Assumption}
\theoremstyle{definition}
\newtheorem{rem}{Remark}

\makeatother

\begin{document}
\title{Regression Discontinuity Design with Potentially Many Covariates\thanks{This paper is a developed version of the previous manuscript (https://sticerd.lse.ac.uk/dps/em/em601.pdf)
inspired by the discussion with Matias Cattaneo. We are grateful to
Matias Cattaneo and Sebastian Calonico for helpful comments and discussions.
This research was supported by Grants-in-Aid for Scientific Research
20K01598 from the Japan Society for the Promotion of Science (Arai)
and the ERC Consolidator Grant (SNP 615882) (Otsu). Financial support
from the Center for National Competitiveness in the Institute of Economic
Research of Seoul National University and the Ministry of Education
of the Republic of Korea and the National Research Foundation of Korea
(NRF-2018S1A5A2A01033487) is gratefully acknowledged (Seo).}}
\author{Yoichi Arai,\thanks{School of Social Sciences, Waseda University, 1-6-1 Nishiwaseda, Shinjuku-ku,
Tokyo 169-8050, Japan. Email: yarai@waseda.jp}\ \ Taisuke Otsu\thanks{Department of Economics, London School of Economics, Houghton Street,
London, WC2A 2AE, UK. Email: t.otsu@lse.ac.uk}\ \ and Myung Hwan Seo\thanks{Department of Economics, Seoul National University, 1 Gwankro Gwanakgu,
Seoul, 08826, Korea. Email: myunghseo@snu.ac.kr}}
\maketitle
\begin{abstract}
This paper studies the case of possibly high-dimensional covariates
in the regression discontinuity design (RDD) analysis. In particular,
we propose estimation and inference methods for the RDD models with
covariate selection which perform stably regardless of the number
of covariates. The proposed methods combine a localization approach
using kernel weights with $\ell_{1}$-penalization to handle high-dimensional
covariates. We provide theoretical and numerical results which illustrate
the usefulness of the proposed methods. Theoretically, we present
risk and coverage properties for our point estimation and inference
methods, respectively. Under certain special cases, the proposed estimator
becomes more efficient than the conventional covariate adjusted estimator
at the cost of an additional sparsity condition. Numerically, our
simulation experiments and empirical example show the robust behaviors
of the proposed methods to the number of covariates in terms of bias
and variance for point estimation and coverage probability and interval
length for inference.
\end{abstract}

\section{Introduction}

In causal or treatment effect analysis, discontinuities in regression
functions induced by an assignment variable can provide useful information
to identify certain causal effects. The regression discontinuity design
(RDD) has been widely applied in observational studies to identify
the average treatment effect at the discontinuity point. For the RDD,
the causal parameters of interest are identified by some contrasts
of the left and right limits of the conditional mean functions. See
e.g. Imbens and Lemieux (2008), Cattaneo, Titiunik and Vazquez-Bare
(2020), an edited volume by Cattaneo and Escanciano (2017), and references
therein.

In the growing literature on the RDD analysis, this paper focuses
on the RDDs where covariates are included in the estimation, which
are extensively studied by Calonico, Cattaneo, Farrell and Titiunik
(2019) (hereafter, CCFT). See also Frölich and Huber (2019) for an
alternative estimation method based on kernel smoothing after localization
around the cutoff. In practice, researchers often augment the regression
models for the RDD analysis with various additional predetermined
covariates such as demographic or socioeconomic characteristics for
data units. For several RDD estimators using covariates based on local
polynomial regression methods, CCFT investigated the MSE expansion,
asymptotic efficiency, and data-driven bandwidth selection methods.
Furthermore, CCFT developed asymptotic distributional approximations
for those estimators and proposed valid inference procedures by constructing
bias and variance estimators with covariate adjustment. These results
may be considered as extensions of the analyses in Calonico, Cattaneo
and Titiunik (2014) (hereafter, CCT) combined with robust bias correction
methods in Calonico, Cattaneo and Farrell (2018, 2020) to incorporate
covariates in the RDD analysis. See also Calonico, Cattaneo, Farrell
and Titiunik (2017) for a statistical package on these methods.

In randomized controlled trials, regression adjustment using covariates
is a common practice since it is always helpful to improve asymptotic
efficiency of the causal effect estimator as far as a full set of
treatment-covariate interactions is included (Lin, 2013). Also a recent
paper by Lei and Ding (2021) proposed a bias correction method for
the regression adjustment estimator with a diverging number of covariates.
On the other hand, in the RDD analysis, which is a quasi-experiment
setup, the efficiency gain by introducing covariates is not necessarily
guaranteed, and CCFT provided a concrete guideline by clarifying the
conditions to achieve consistency and efficiency gain for the covariate
adjusted RDD estimator. Typically the efficiency improves when the
projection coefficients of the covariates on the outcome are equal
for both control and treatment groups. Since practitioners also commonly
incorporate covariates for the RDD analysis, CCFT's guideline has
a large impact in applied research. When we use covariates, it is
common to employ their transformations and interactions, and the number
of these terms can be pretty large. This paper adds a further guideline
for practitioners who also face a large number of covariates. To begin
with, the (weighted) OLS estimation in CCFT is not applicable when
the number of covariates is larger than the sample size. Also, in
the above scenario for efficiency improvement, it is beneficial to
augment CCFT's procedure with covariate selection by high-dimensional
statistical methods particularly when the regression coefficients
for the conditional mean function satisfy certain sparsity.

For point estimation on the causal effect parameter identified by
the RDD, we consider the Lasso estimator and its post-selection estimator
based on the local linear regression (i.e., eq. (2) of CCFT). The
combination of localization using kernel weights and $\ell_{1}$-penalization
to deal with high-dimensional covariates is particularly relevant
for the RDD analysis, where the effective sample size would be typically
small due to the localization so that the effect of dimensionality
of covariates becomes severer. Theoretically, we derive the $\ell_{1}$-risk
properties of our local Lasso estimator and its post-selection version.
Practically, based on our simulation study, we recommend the CCFT
estimator after selecting covariates by the $\ell_{1}$-penalization
even for a relatively small number of covariates, which exhibits desirable
MSE properties and stability across different setups.

For inference, we propose to select covariates with the local Lasso.
We show that the inference based on the selected covariates can be
implemented in the same manner as in CCFT. We also show that when
the effect of the additional covariates on the potential outcomes
with or without treatment is invariant, our approach can lead to improved
efficiency at the cost of an additional sparsity condition. This sparsity
condition is trivially satisfied when the set of active covariates
is unknown but fixed. Our simulation results demonstrate that our
post-selection confidence interval exhibits robust performances in
terms of both coverages and lengths, even for a relatively small number
of covariates.

This paper also contributes to the large literature on high-dimensional
methods in econometrics and statistics (see, e.g., Bühlmann and van
de Geer, 2011, and Belloni \emph{et al.}, 2018, for an overview) by
combining the kernel localization with $\ell_{1}$-penalization to
handle high-dimensional covariates. Our inference problem can be formulated
as the one for low-dimensional parameters in high-dimensional models.
In statistics literature, many papers investigated this issue, such
as Belloni, Chernozhukov and Hansen (2014), van de Geer, \emph{et
al.} (2014), and Zhang and Zhang (2014). However, these approaches
are not directly applicable to the RDD context because the current
problem concerns the inference on a jump in a nonparametric regression
model.\footnote{A recent paper by Krei\ss\ and Rothe (2023) investigates a similar
estimator to ours, and discusses an inference method based on the
approach by Armstrong and Kolesár (2018).}

This paper is organized as follows. Section \ref{sub:setup} introduces
our basic setup and local Lasso estimator, and presents the $\ell_{1}$-risk
properties. In Section \ref{sub:inf}, we discuss the validity of
CCFT's inference after selecting covariates by our Lasso procedure.
Section \ref{sec:dis} provides discussions on some extensions. A
step-by-step procedure for implementation of our method is described
in Section \ref{sec:rec}. To illustrate the proposed method, Section
\ref{sec:sim} conducts a simulation study, and Section \ref{sec:emp}
presents an empirical example based on the Head Start data.

\section{Main result\label{sec:main}}

\subsection{Setup and local Lasso estimator for covariate selection\label{sub:setup}}

In this subsection, we present our basic setup and introduce the local
Lasso estimator for the RDD with possibly high-dimensional covariates.
For each unit $i=1,\ldots,n$, we observe an indicator variable $T_{i}$
for a treatment ($T_{i}=1$ if treated and $T_{i}=0$ otherwise),
and outcome $Y_{i}=Y_{i}(0)\cdot(1-T_{i})+Y_{i}(1)\cdot T_{i}$, where
$Y_{i}(0)$ and $Y_{i}(1)$ are potential outcomes for $T_{i}=0$
and $T_{i}=1$, respectively. Note that we cannot observe $Y_{i}(0)$
and $Y_{i}(1)$ simultaneously. Our purpose is to make inference on
the causal effect of the treatment, or more specifically, some distributional
aspects of the difference of the potential outcomes $Y_{i}(1)-Y_{i}(0)$.
The RDD analysis focuses on the case where the treatment assignment
$T_{i}$ is completely or partly determined by some observable covariate
$X_{i}$, called the running variable. For example, to study the effect
of class size on pupils' achievements, it is reasonable to consider
the following setup: the unit $i$ is school, $Y_{i}$ is an average
exam score, $T_{i}$ is an indicator variable for the class size ($T_{i}=0$
for one class and $T_{i}=1$ for two classes), and $X_{i}$ is the
number of enrollments. For more examples, see e.g. Imbens and Lemieux
(2008), Cattaneo, Titiunik and Vazquez-Bare (2020), Cattaneo and Escanciano
(2017), and references therein.

Depending on the assignment rule for $T_{i}$ based on $X_{i}$, we
have two cases, called the sharp and fuzzy RDDs. In this section,
we focus on the sharp RDD and discuss the fuzzy RDD in Section \ref{sub:fuzzy}.
In the sharp RDD, the treatment is deterministically assigned as $T_{i}=\mathbb{I}\{X_{i}\geq\bar{x}\}$,
where $\mathbb{I}\{\cdot\}$ is the indicator function and $\bar{x}$
is a known discontinuity (cutoff) point. Throughout the paper, we
normalize $\bar{x}=0$ to simplify the presentation. A parameter of
interest, in this case, is the average causal effect at the discontinuity
point:
\begin{equation}
\tau=\mathbb{E}[Y_{i}(1)-Y_{i}(0)|X_{i}=0].\label{eq:tau}
\end{equation}
Since the difference $Y_{i}(1)-Y_{i}(0)$ is unobservable, we need
a tractable representation of $\tau$ in terms of quantities that
can be estimated by data. If the conditional mean functions $\mathbb{E}[Y_{i}(1)|X_{i}=x]$
and $\mathbb{E}[Y_{i}(0)|X_{i}=x]$ are continuous at the cutoff point
$x=0$, then the average causal effect $\tau$ can be identified as
a contrast of the left and right limits of the conditional mean $\mathbb{E}[Y_{i}|X_{i}=x]$
at $x=0$, that is
\begin{equation}
\tau=\lim_{x\downarrow0}\mathbb{E}[Y_{i}|X_{i}=x]-\lim_{x\uparrow0}\mathbb{E}[Y_{i}|X_{i}=x].\label{eq:tau1}
\end{equation}

As argued in CCFT, it is usually the case that practitioners have
access to additional covariates (denoted by $Z_{i}\in\mathbb{R}^{p}$)
and augment their empirical models with $Z_{i}$ to estimate the causal
effect $\tau$ of interest. This practically relevant setup is extensively
studied in CCFT for the case where $Z_{i}$ is low-dimensional. In
this paper, we consider the case of possibly high-dimensional $Z_{i}$,
and propose a new point estimation method for $\tau$ and an adjustment
of CCFT's inference method.

We examine the case where the additional covariates $Z_{i}$ are predetermined
in the sense that $Z_{i}=Z_{i}(0)\cdot(1-T_{i})+Z_{i}(1)\cdot T_{i}$
but $Z_{i}(0)=_{d}Z_{i}(1)$ for the potential covariates $Z_{i}(0)$
and $Z_{i}(1)$ for $T_{i}=0$ and $T_{i}=1$, respectively. Motivated
by CCFT's recommended model (in their eq. (2)), we propose the local
Lasso estimator $\hat{\theta}=(\hat{\alpha},\hat{\tau},\hat{\beta}_{-},\hat{\beta}_{+},\hat{\gamma}^{\prime})^{\prime}$
that solves
\begin{equation}
\min_{\alpha,\tau,\beta_{-},\beta_{+},\gamma}\frac{1}{nb_{n}}\sum_{i=1}^{n}K\left(\frac{X_{i}}{b_{n}}\right)(Y_{i}-\alpha-T_{i}\tau-X_{i}\beta_{-}-T_{i}X_{i}\beta_{+}-Z_{i}^{\prime}\gamma)^{2}+\lambda_{n}|\gamma|_{1},\label{eq:lasso}
\end{equation}
where $|\gamma|_{1}=\sum_{j=1}^{p}|\gamma_{j}|$ is the $\ell_{1}$-norm
of $\gamma$, $\gamma_{j}$ means the $j$-th element of $\gamma$,
$K(\cdot)$ is a kernel function, $b_{n}$ is a bandwidth, and $\lambda_{n}$
is a penalty level. Popular choices for $K(\cdot)$ are the uniform
and triangular kernels supported on $[-b_{n},b_{n}]$. Based on (\ref{eq:lasso}),
our point estimator for $\tau$ is given by $\hat{\tau}$.

Our preliminary simulation results suggest that the local Lasso estimator
for $\tau$ is somewhat biased in finite samples. Therefore, our recommendation
for point estimation is to employ a post-selection method. Let $\hat{S}=\{j:|\hat{\gamma}_{j}|\ge\zeta_{n}\}$
for a non-negative sequence $\{\zeta_{n}\}$, and $Z_{\hat{S},i}$
be a subvector of $Z_{i}$ selected by $\hat{S}$. Then the local
post-Lasso estimator $\bar{\theta}=(\bar{\alpha},\bar{\tau},\bar{\beta}_{-},\bar{\beta}_{+},\bar{\gamma}_{\hat{S}}^{\prime})^{\prime}$
is defined as a solution of the local least square:
\begin{equation}
\min_{\alpha,\tau,\beta_{-},\beta_{+},t}\frac{1}{nh_{n}}\sum_{i=1}^{n}K\left(\frac{X_{i}}{h_{n}}\right)(Y_{i}-\alpha-T_{i}\tau-X_{i}\beta_{-}-T_{i}X_{i}\beta_{+}-Z_{\hat{S},i}^{\prime}t)^{2},\label{eq:post-lasso}
\end{equation}
where $h_{n}$ is another bandwidth, and the estimator for $\tau$
is given by $\bar{\tau}$.

Several points are worthy of remark for this estimator. First, without
the $\ell_{1}$-penalization, our estimator reduces to the local linear-type
estimator recommended by CCFT's eq. (2). Therefore, the proposed estimator
is a natural generalization of CCFT's when the dimension of $Z_{i}$
is high. Second, without the kernel weights for localization, our
estimator in (\ref{eq:lasso}) reduces to the conventional Lasso estimator.
However, since our parameter of interest $\tau$ is identified as
a local object in (\ref{eq:tau1}), it is crucial to introduce such
localization to avoid misspecification bias of the conditional mean
functions. Third, it is often the case that the kernel function $K(\cdot)$
has bounded support. In this case, the effective sample size would
be typically of orders $nb_{n}$ and $nh_{n}$. Thus even if the dimension
of $Z_{i}$ is relatively small compared to the original sample size
$n$, the $\ell_{1}$-penalization would be useful especially for
small values of $b_{n}$ and $h_{n}$. Finally, the trimming term
$\zeta_{n}$ to obtain the set $\hat{S}$ is introduced to stabilize
numerical results (see (\ref{eq:zeta}) below for our recommended
choice based on simulation studies), and theoretically we may set
as $\zeta_{n}=0$.

We now present risk properties of the local Lasso estimators $\hat{\theta}$
and $\bar{\theta}$. Let $G_{i}=(1,T_{i},X_{i},T_{i}X_{i},Z_{i}^{\prime})^{\prime}$
be the vector of regressors in (\ref{eq:lasso}), $G_{i,j}$ be the
$j$-th element of $G_{i}$, and $\Theta_{n}=\arg\min_{\theta}\mathbb{E}[K(X_{i}/b_{n})(Y_{i}-G_{i}^{\prime}\theta)^{2}]$
be an argmin set. We impose the following assumptions.

\begin{asm}\label{asm:point-est} There exists a sequence $\theta_{n}^{*}\in\Theta_{n}$
that satisfies the following conditions.
\begin{enumerate}
\item Let $\epsilon_{i}=\sqrt{K(X_{i}/b_{n})}(Y_{i}-G_{i}^{\prime}\theta_{n}^{*})$.
There exists some $C\in(0,\infty)$ such that
\[
\mathbb{E}[|K(X_{i}/b_{n})G_{i,j}\epsilon_{i}|^{m}]\leq b_{n}m!C^{m-2}/2,
\]
for all $j=1,\ldots,p$ and $m=2,3,\ldots$.
\item Let $\delta_{A}$ be the subvector of $\delta$ for an index set $A$,
$S^{*}=\{j:\theta_{n,j}^{*}\neq0\}$, and $(S^{*})^{c}=\{j:\theta_{n,j}^{*}=0\}$,
where $\theta_{n,j}^{*}$ is the $j$-th element of $\theta_{n}^{*}$.
There exists some $\phi^{*}\in(0,\infty)$ such that
\[
\frac{s^{*}}{|\delta_{S^{*}}|_{1}^{2}}\cdot\min_{\delta:|\delta_{(S^{*})^{c}}|_{1}\leq3|\delta_{S^{*}}|_{1}}\delta^{\prime}\left(\frac{1}{nb_{n}}\sum_{i=1}^{n}K\left(\frac{X_{i}}{b_{n}}\right)G_{i}G_{i}^{\prime}\right)\delta\geq(\phi^{*})^{2},
\]
with probability approaching one, where $s^{*}=|S^{*}|$.
\end{enumerate}
\end{asm}

\begin{asm} \label{asm:kb} $K:\mathbb{R\to\mathbb{R}}$ is a bounded
and symmetric second-order kernel function which is continuous with
a compact support. The bandwidth $b_{n}$ is a positive sequence satisfying
$b_{n}\to0$ and $nb_{n}\to\infty$ as $n\to\infty$. \end{asm}

Assumption \ref{asm:point-est} defines $\theta_{n}^{*}$ as an approximate
linear predictor or the linear projection on the set of included variables
in the index set $S^{*}$ since $\mathbb{E}[K(X_{i}/b_{n})G_{i,j}\epsilon_{i}]=0$
for all $j\in S^{*}$. This assumption is general enough to cover
the setup in CCFT, which assumes $p$ is fixed. In RDD analyses, it
is common to introduce many generated covariates, such as transformations
of initial covariates like polynomials, interactions, and various
basis functions, without knowing which of them are relevant a priori.\footnote{To motivate the use of generated covariates, it is insightful to note
that the asymptotic variance of CCFT's RDD estimator is proportional
to $\mathrm{Var}(\{(Y_{i}(1)-Y_{i}(0))-(Z_{i}(1)-Z_{i}(0))^{\prime}\gamma\}^{2}|X_{i}=0)$
for some $\gamma$, which is considered as the (conditional) variance
of the linear projection error. Although CCFT considered linear projection
due to the constraint on dimensionality, it is clear that the asymptotic
variance is minimized by employing the conditional expectation $\mathbb{E}[Y_{i}(1)-Y_{i}(0)|Z_{i}(1)-Z_{i}(0),X_{i}=0]$
instead of the linear projection $(Z_{i}(1)-Z_{i}(0))^{\prime}\gamma$.
Therefore, it is natural to extend CCFT's approach to high-dimensional
settings by employing generated covariates or series approximation
for the conditional mean.}

Although it is beyond the scope of this paper, the definition of $\theta_{n}^{*}$
could be modified to be an approximate minimizer which does not belong
to $\Theta_{n}$ but gets closer to it at some suitable rate. Since
it complicates the exposition and derivation as in Krei\ss\ and
Rothe (2023) or Belloni, Chernozhukov and Hansen (2014), we maintain
this exact sparsity assumption. For example, such an extension for
approximate sparsity will be useful to allow the situation where the
conditional mean satisfies \emph{$E[Y|X,Z]=E[Y|X,Z_{\mathcal{S}}]$}
for some sparse set $\mathcal{S}$ but the conditional mean function
$E[Y|X,Z_{\mathcal{S}}]$ is nonlinear in $Z_{\mathcal{S}}$ so that
the exact sparsity assumption typically fails.

Assumption \ref{asm:point-est} (1) contains a set of moment conditions,
which are introduced to verify a local-type Bernstein inequality in
Lemma \ref{lem:Bernstein}. The extra factor $b_{n}$ is due to the
presence of the kernel weight $K(X_{i}/b_{n})$. Assumption \ref{asm:point-est}
(2) is a localized version of the compatibility condition. A sufficient
condition for this is the so-called restricted eigenvalue condition.
More specifically, $\min_{\beta:|\beta|_{0}\leq s^{*}}\frac{1}{nb_{n}}\sum_{i=1}^{n}K\left(\frac{X_{i}}{b_{n}}\right)\frac{\beta^{\prime}G_{i}G_{i}^{\prime}\beta}{\beta^{\prime}\beta}$,
where $|\beta|_{0}$ denotes the cardinality of $\beta$, provides
a lower bound for the compatibility constant $(\phi^{*})^{2}$ (see,
e.g., Section 6.13 of Bühlmann and van der Geer, 2011). If CCFT's
method is feasible with each subset of covariates of dimension $2s^{*}$,
then the restricted eigenvalue condition is indeed satisfied. Assumption
\ref{asm:kb} contains assumptions on the kernel $K$ and bandwidth
$b_{n}$, which are standard in the literature of nonparametric methods.
Note that since $\theta_{n}^{*}$ and $S^{*}$ depend on $b_{n}$,
Assumption \ref{asm:point-est} should be satisfied along each sequence
$\{b_{n}\}$. Also, the deviation bounds on the prediction and estimation
errors of $\hat{\theta}$ will be given as functions of $s^{*}$.
While a precise condition on $s^{*}$ is hard to specify and depends
on the sampling distribution, it would be typically smaller order
than $\sqrt{nb_{n}}$ to satisfy the compatibility condition in Assumption
\ref{asm:point-est} (2).

Let $\hat{\gamma}_{\hat{S}}$ be the subvector of $\hat{\gamma}$
selected by $\hat{S}$, $\hat{\theta}_{\hat{S}}=(\hat{\alpha},\hat{\tau},\hat{\beta}_{-},\hat{\beta}_{+},\hat{\gamma}_{\hat{S}}^{\prime})^{\prime}$,
$Z_{\hat{S},i}$ be the subvector of $Z_{i}$ selected by $\hat{S}$,
$G_{\hat{S},i}=(1,T_{i},X_{i},T_{i}X_{i},Z_{\hat{S},i}^{\prime})^{\prime}$,
and $m_{n}=\lambda_{\min}\left(\frac{1}{nh_{n}}\sum_{i=1}^{n}K(X_{i}/h_{n})G_{\hat{S},i}G_{\hat{S},i}^{\prime}\right)^{-1}$,
where $\lambda_{\min}(A)$ means the minimum eigenvalue of a matrix
$A$. Also let $\hat{S}_{1}=\{i:0<|\hat{\gamma}_{i}|<\zeta_{n}\}$,
$S_{n}=\hat{S}\cup\hat{S}_{1}=\{i:\hat{\gamma}_{i}\neq0\}$, $Z_{\hat{S}_{1},i}$
be the subvector of $Z_{i}$ selected by $\hat{S}_{1}$, $G_{\hat{S}_{1},i}=(1,T_{i},X_{i},T_{i}X_{i},Z_{\hat{S}_{1},i}^{\prime})^{\prime}$,
and $m_{1n}=\lambda_{\max}\left(\frac{1}{nh_{n}}\sum_{i=1}^{n}K(X_{i}/h_{n})G_{\hat{S}_{1},i}G_{\hat{S}_{1},i}^{\prime}\right)$,
where $\lambda_{\max}(A)$ means the maximum eigenvalue of a matrix
$A$. The $\ell_{1}$-risk properties of the local Lasso and post-Lasso
estimators (for the case of $h_{n}=b_{n}$) are obtained as follows. 

\begin{thm}\label{thm:point-est} Suppose $\lambda_{n}^{-1}\sqrt{\log p/(nb_{n})}\to0$.
\begin{description}
\item [{(i)}] Under Assumptions \ref{asm:point-est}-\ref{asm:kb}, it
holds
\begin{equation}
|\hat{\theta}-\theta_{n}^{\ast}|_{1}\leq C\frac{\lambda_{n}s^{*}}{\phi^{*2}},\label{eq:l1b}
\end{equation}
for some $C\in(0,\infty)$ with probability approaching one.
\item [{(ii)}] Under Assumptions \ref{asm:point-est}-\ref{asm:kb} and
$h_{n}=b_{n}$, it holds
\begin{equation}
|\bar{\theta}_{S_{n}}-\hat{\theta}_{S_{n}}|_{1}\leq(m_{n}|S_{n}|\lambda_{n})\vee(|\hat{S}_{1}|\zeta_{n})\vee(m_{1n}|\hat{S}_{1}|\zeta_{n}^{2}/\lambda_{n}).\label{eq:l1c}
\end{equation}
\end{description}
\end{thm}

The proof of this theorem is presented in Appendix \ref{sub:pf1}.
This theorem characterizes the risk properties of the estimators $\hat{\theta}$
and $\bar{\theta}$ around $\theta_{n}^{*}$. The risk bound of $\hat{\theta}$
depends on the tuning parameter $\lambda_{n}$, the number of non-zero
coefficients $s^{*}$, and the compatibility constant $\phi^{*}$.
Note that the decay rate of $\lambda_{n}$ is bounded from below by
$\sqrt{\log p/(nb_{n})}$. Thus, the risk bound of $\hat{\theta}$
gets worse as the number of covariates $p$ increases or the effective
sample size $nb_{n}$ due to the kernel localization decreases. The
result (\ref{eq:l1c}) for the post-selection estimator $\bar{\theta}$
shows that the deviation from the original Lasso estimator $\hat{\theta}$
is small when tuning parameter $\lambda_{n}$ or the number of selected
covariates $|S_{n}|$ is small, or the minimum eigenvalue of $\frac{1}{nh_{n}}\sum_{i=1}^{n}K(X_{i}/b_{n})G_{\hat{S},i}G_{\hat{S},i}^{\prime}$
is large. If the trimming parameter $\zeta_{n}$ is of similar magnitude
of $\lambda_{n}$, then the three terms in the bounds are of similar
magnitude. If $\zeta_{n}$ is smaller order of magnitude than $\lambda_{n}$,
then the first term will dominate. We suggest some practical choice
of the trimming term $\zeta_{n}$ in Section \ref{sec:sim} based
on our simulation studies. 

The above theorem is on estimation of the coefficients of the best
linear predictor $\theta_{n}^{*}$ defined in Assumption \ref{asm:point-est}
(1). Additionally suppose that the assumptions of Lemma 1 of CCFT
hold true, and the covariates $Z_{i}$ are predetermined. Then we
can guarantee that the second element of $\theta_{n}^{*}$ coincides
with the average causal effect $\tau$ in (\ref{eq:tau}) so that
Theorem \ref{thm:point-est} provides the conditions for the consistency
and convergence rate of $\hat{\tau}$ to $\tau$. If $Z_{i}$ are
not predetermined (i.e., $Z_{i}(0)\neq_{d}Z_{i}(1)$), then $\hat{\tau}$
typically converges to $\tau$ minus some bias component, which is
obtained as a limit of CCFT's bias term in their Lemma 1.

Our estimators and above theorem can be extended to other regression
models that contain the covariates $\{T_{i}Z_{i},(1-T_{i})Z_{i}\}$,
$(Z_{i}-\bar{Z})$, or $\{T_{i}(Z_{i}-\bar{Z}),(1-T_{i})(Z_{i}-\bar{Z})\}$
as in CCFT. However, as shown in Lemma 1 of CCFT, such estimators
require more stringent conditions to guarantee the consistency for
$\tau$. Furthermore, the local Lasso regression (\ref{eq:lasso})
can be extended to incorporate polynomials of $X_{i}$ and $T_{i}X_{i}$
even though this paper focuses on the local linear model.

Finally, we discuss the choices of the localization bandwidths $b_{n}$
and $h_{n}$ and regularization parameter $\lambda_{n}$. We can use
the MSE-optimal bandwidth based on the suggestion by CCFT and the
regularization parameter $\lambda_{n}$ using cross-validation by
Friedman, Hastie and Tibshirani (2010) or a data-driven choice by
Belloni, Chernozhukov and Hansen (2014) among others. See Theorem
\ref{thm:t} in the next subsection for their justification, and Section
\ref{sec:rec} for a detail on our practical recommendation.

\subsection{Inference\label{sub:inf}}

We next consider interval estimation and hypothesis testing on the
average causal effect $\tau$. For finite or low-dimensional $Z_{i}$,
we recommend to use CCFT's bias corrected inference method. This subsection
argues that we can still apply CCFT's inference procedure for high-dimensional
$Z_{i}$, provided that CCFT's conditions remain valid for $S^{*}$
and the subvector $\theta_{n,S^{*}}^{*}$ of $\theta_{n}^{*}$ selected
by $S^{*}$.

In this subsection, we specify the tuning constant $\zeta_{n}$ to
obtain $\hat{S}=\{j:|\hat{\gamma}_{j}|\ge\zeta_{n}\}$ as
\begin{equation}
\zeta_{n}=\lambda_{n}\varrho_{n}\sum_{j=1}^{p}\mathbb{I}\{\hat{\gamma}_{j}\neq0\},\label{eq:zeta}
\end{equation}
where we set $\varrho_{n}=\log\log\log n$. This choice of $\varrho_{n}$
is based on the simulation experiments in Section \ref{sec:sim},
and it is not shown to be optimal but works reasonably well. Based
on the selected covariates by $\hat{S}$ with $\zeta_{n}$ in (\ref{eq:zeta}),
we apply CCFT's bias corrected t-ratio to conduct statistical inference
on the causal effect parameter $\tau$.

Consider the local post-Lasso estimator $\bar{\tau}$ defined by (\ref{eq:post-lasso}).
As shown in Appendix \ref{sub:pf2}, the dominant term of $\bar{\tau}$
can be characterized as
\begin{equation}
\bar{\tau}=e_{2}^{\prime}\left(\frac{1}{nh_{n}}\sum_{i=1}^{n}K\left(\frac{X_{i}}{h_{n}}\right)G_{1i}G_{1i}^{\prime}\right)^{-1}\frac{1}{nh_{n}}\sum_{i=1}^{n}K\left(\frac{X_{i}}{h_{n}}\right)G_{1i}\xi_{i}+o_{p}((nh_{n})^{-1/2}),\label{eq:lin}
\end{equation}
where $e_{2}=(0,1,0,0)^{\prime}$, $G_{1i}=(1,T_{i},X_{i},T_{i}X_{i})^{\prime}$,
$\xi_{i}=T_{i}\xi_{i}(1)+(1-T_{i})\xi_{i}(0)$ with $\xi_{i}(t)=Y_{i}(t)-Z_{i}(t)^{\prime}\gamma_{Y}$,
and
\begin{equation}
\gamma_{Y}=\arg\min_{\gamma:\gamma_{j}=0\text{ for }j\notin S^{*}}\mathbb{E}[(\tilde{Y}-Z^{\prime}\gamma)^{2}|X=0],\label{eq:gamY}
\end{equation}
with $\tilde{Y}=Y(1)-Y(0)-\mathbb{E}[Y(1)-Y(0)|X=0]$. Indeed, the
asymptotic linear form in (\ref{eq:lin}) is analogous to the one
derived for the case of fixed dimensional $Z_{i}$ in CCFT (except
that $\gamma_{Y,j}=0$ for $j\notin S^{*}$). Therefore, the pre-asymptotic
bias and variance of $\bar{\tau}$ can be analogously written as 
\begin{eqnarray}
\mathcal{B} & = & \frac{1}{2}e_{1}^{\prime}\Gamma_{-}^{-1}\vartheta_{-}(q^{\prime}\mu_{-}^{(2)})+\frac{1}{2}e_{1}^{\prime}\Gamma_{+}^{-1}\vartheta_{+}(q^{\prime}\mu_{+}^{(2)}),\nonumber \\
\mathcal{V} & = & (q\otimes P_{-}^{\prime}e_{1})^{\prime}\Sigma_{-}(q\otimes P_{-}^{\prime}e_{1})+(q\otimes P_{+}^{\prime}e_{1})^{\prime}\Sigma_{+}(q\otimes P_{+}^{\prime}e_{1}),\label{eq:V}
\end{eqnarray}
respectively, where $e_{1}=(1,0)^{\prime}$, $R=\left[\begin{array}{ccc}
1 & \ldots & 1\\
X_{1}/h_{n} & \ldots & X_{n}/h_{n}
\end{array}\right]^{\prime}$, $q=(1,-\gamma^{*\prime})^{\prime}$, and
\begin{eqnarray}
K_{-} & = & h_{n}^{-1}\mathrm{diag}(\mathbb{I}\{X_{1}<0\}K(X_{1}/h_{n}),\ldots,\mathbb{I}\{X_{n}<0\}K(X_{n}/h_{n})),\nonumber \\
K_{+} & = & h_{n}^{-1}\mathrm{diag}(\mathbb{I}\{X_{1}\ge0\}K(X_{1}/h_{n}),\ldots,\mathbb{I}\{X_{n}\ge0\}K(X_{n}/h_{n})),\nonumber \\
\Gamma_{-} & = & n^{-1}R^{\prime}K_{-}R,\qquad\Gamma_{+}=n^{-1}R^{\prime}K_{+}R,\nonumber \\
\vartheta_{-} & = & n^{-1}R^{\prime}K_{-}[X_{1}^{2}/h_{n}^{2},\ldots,X_{n}^{2}/h_{n}^{2}]^{\prime},\qquad\vartheta_{+}=n^{-1}R^{\prime}K_{+}[X_{1}^{2}/h_{n}^{2},\ldots,X_{n}^{2}/h_{n}^{2}]^{\prime},\nonumber \\
\mu_{-}^{(2)} & = & \left[\left.\frac{\partial^{2}\mathbb{E}[Y_{i}(0)|X_{i}=x]}{\partial x^{2}}\right|_{x=0},\left.\frac{\partial^{2}\mathbb{E}[Z_{i}^{*}(0)^{\prime}|X_{i}=x]}{\partial x^{2}}\right|_{x=0}\right]^{\prime},\nonumber \\
\mu_{+}^{(2)} & = & \left[\left.\frac{\partial^{2}\mathbb{E}[Y_{i}(1)|X_{i}=x]}{\partial x^{2}}\right|_{x=0},\left.\frac{\partial^{2}\mathbb{E}[Z_{i}^{*}(1)^{\prime}|X_{i}=x]}{\partial x^{2}}\right|_{x=0}\right]^{\prime},\nonumber \\
P_{-} & = & \sqrt{\frac{h}{n}}\Gamma_{-}^{-1}R^{\prime}K_{-},\qquad P_{+}=\sqrt{\frac{h}{n}}\Gamma_{+}^{-1}R^{\prime}K_{+},\nonumber \\
\Sigma_{-} & = & \mathrm{Var}(\mathrm{vec}(\mathbf{Y}(0),\mathbf{Z}^{*}(0))|\mathbf{X}=0),\qquad\Sigma_{+}=\mathrm{Var}(\mathrm{vec}(\mathbf{Y}(1),\mathbf{Z}^{*}(1))|\mathbf{X}=0),\label{eq:Vnote}
\end{eqnarray}
with $\mathbf{Y}(0)=(Y_{1}(0),\ldots,Y_{n}(0))^{\prime}$, $\mathbf{Y}(1)=(Y_{1}(1),\ldots,Y_{n}(1))^{\prime}$,
$\mathbf{X}=(X_{1},\ldots,X_{n})^{\prime}$, $\mathbf{Z}^{*}(0)=(Z_{S^{*},1}(0),\ldots,Z_{S^{*},n}(0))^{\prime}$,
and $\mathbf{Z}^{*}(1)=(Z_{S^{*},1}(1),\ldots,Z_{S^{*},n}(1))^{\prime}$.

By estimating the unknown components, the pre-asymptotic bias and
variance can be estimated as
\begin{eqnarray*}
\bar{\mathcal{B}} & = & \frac{1}{2}e_{1}^{\prime}\Gamma_{-}^{-1}\vartheta_{-}(\bar{q}^{\prime}\bar{\mu}_{-}^{(2)})+\frac{1}{2}e_{1}^{\prime}\Gamma_{+}^{-1}\vartheta_{+}(\bar{q}^{\prime}\bar{\mu}_{+}^{(2)}),\\
\bar{\mathcal{V}} & = & (\bar{q}\otimes P_{-}^{\prime}e_{1})^{\prime}\bar{\Sigma}_{-}(\bar{q}\otimes P_{-}^{\prime}e_{1})+(\bar{q}\otimes P_{+}^{\prime}e_{1})^{\prime}\bar{\Sigma}_{+}(\bar{q}\otimes P_{+}^{\prime}e_{1}),
\end{eqnarray*}
respectively, where $\bar{q}^{\prime}=(1,-\bar{\gamma}_{\hat{S}}^{\prime})^{\prime}$,
$\bar{\mu}_{-}^{(2)}$ and $\bar{\mu}_{+}^{(2)}$ are local polynomial
estimators of $\mu_{-}^{(2)}$ and $\mu_{+}^{(2)}$ for the elements
corresponding to $Z_{\hat{S},i}$, respectively, and $\bar{\Sigma}_{+}$
and $\bar{\Sigma}_{-}$ are conditional variance estimators of $\Sigma_{-}$
and $\Sigma_{+}$, respectively, such as the nearest neighborhood
or plug-in estimator in Section 7.9 of CCFT's supplement. Based on
these estimators, the t-ratio for $\tau$ is obtained as
\begin{equation}
T_{\tau}=\frac{\bar{\tau}-h_{n}^{2}\bar{\mathcal{B}}-\tau}{\sqrt{(nh_{n})^{-1}\bar{\mathcal{V}}}},\label{eq:t}
\end{equation}
which is exactly the same as the t-ratio in Theorem 2 of CCFT but
using the selected covariates $Z_{\hat{S},i}$. By extending the theoretical
developments in CCFT, we obtain the following result.

\begin{thm} \label{thm:t} Suppose Assumptions \ref{asm:point-est}-\ref{asm:kb}
hold true. Suppose for all $x$ in a neighborhood of $0$ and $t=0,1$,
the density of $X_{i}$ is continuous and bounded away from zero,
$\mathbb{E}[(Y_{i}(t),Z_{i}(t)^{\prime})|X_{i}=x]$ is three times
continuously differentiable, $\partial^{2}\mathbb{E}[Z_{i}(0)|X_{i}=x]/\partial x^{2}=\partial^{2}\mathbb{E}[Z_{i}(1)|X_{i}=x]/\partial x^{2}$,
$\mathbb{E}[Z_{i}(t)Y_{i}(t)|X_{i}=x]$ is continuously differentiable,
$\mathrm{Var}((Y_{i}(t),Z_{i}(t)^{\prime})|X_{i}=x)$ is continuously
differentiable and invertible, and $\mathbb{E}[|(Y_{i}(t),Z_{i}(t)^{\prime})|^{4}|X_{i}=x]$
is continuous. For all $j,k$ and positive integers $m$, and some
finite $C$, $\mathbb{E}[|K(X_{i}/h_{n})G_{1i,j}Z_{i,k}|^{m}]\leq h_{n}m!C^{m-2}/2$.
Finally, assume $h_{n}\to0$, $nh_{n}\to\infty$, $\lambda_{n}^{-1}\sqrt{\log p/(nb_{n})}\to0$,
and $(\sqrt{\log p}+\sqrt{nh_{n}}h_{n}^{2})(h_{n}^{2}+b_{n}^{2}+\lambda_{n}+\zeta_{n}+\zeta_{n}^{2}/\lambda_{n})s^{*}\to0$.
Then the conditional MSE expansion of the first term in (\ref{eq:lin})
(denoted by $\bar{\tau}_{1}$) is obtained as
\begin{equation}
\mathbb{E}[(\bar{\tau}_{1}-\tau)^{2}|\mathbf{X}]=h_{n}^{4}\mathcal{B}^{2}\{1+o_{p}(1)\}+\frac{1}{nh_{n}}\mathcal{V}.\label{eq:MSE}
\end{equation}
Furthermore, if we additionally assume $\sqrt{\frac{nh_{n}^{5}}{\mathcal{V}}}(\bar{\mathcal{B}}-\mathcal{B})\overset{p}{\to}0$
and $\frac{\bar{\mathcal{V}}}{\mathcal{V}}\overset{p}{\to}1$, then

\begin{eqnarray}
T_{\tau} & \overset{d}{\to} & N(0,1).\label{eq:tau-t}
\end{eqnarray}
 \end{thm}

The results in (\ref{eq:MSE}) and (\ref{eq:tau-t}) are analogous
to CCFT's Theorems 1 and 2, respectively. This theorem theoretically
supports to employ the bias correction and bandwidth selection methods
by CCFT based on the selected covariates $Z_{\hat{S},i}$. See Section
\ref{sec:rec} below for our practical recommendation. The assumption
$\partial^{2}\mathbb{E}[Z_{i}(0)|X_{i}=x]/\partial x^{2}=\partial^{2}\mathbb{E}[Z_{i}(1)|X_{i}=x]/\partial x^{2}$
is natural for predetermined covariates but may be relaxed by introducing
additional regularity conditions (see, Krei\ss\ and Rothe, 2023).
Other assumptions except for the last one are also imposed in CCFT.
The assumption $(\sqrt{\log p}+\sqrt{nh_{n}}h_{n}^{2})(h_{n}^{2}+b_{n}^{2}+\lambda_{n}+\zeta_{n}+\zeta_{n}^{2}/\lambda_{n})s^{*}\to0$
is used to control the remainder term in (\ref{eq:lin}), and can
be considered as a sparsity assumption to restrict the growth rate
of $s^{*}$.\footnote{In the standard Lasso literature, we typically impose $\frac{s^{*}\log p}{\sqrt{n}}\to0$
and the minimal penalty level requirement on $\lambda_{n}$. For comparison,
consider the following standard setting for tuning parameters, where
$b_{n}\sim h_{n}\sim n^{-1/5}$, $\lambda_{n}=a_{n}\sqrt{\log p/(nb_{n})}$
with a slowly diverging $a_{n}$ and $\zeta_{n}=O(\lambda_{n})$.
Then the condition on $s^{*}$ reduces to $\frac{s^{*}a_{n}\log p}{\sqrt{nh_{n}}}\to0$,
which is analogous to the standard case.} Although the conditions $\sqrt{\frac{nh_{n}^{5}}{\mathcal{V}}}(\bar{\mathcal{B}}-\mathcal{B})\overset{p}{\to}0$
and $\frac{\bar{\mathcal{V}}}{\mathcal{V}}\overset{p}{\to}1$ are
high level, these are typically satisfied for the bias and variance
estimators discussed in CCFT.\footnote{Under the assumption $\partial^{2}\mathbb{E}[Z_{i}(0)|X_{i}=x]/\partial x^{2}=\partial^{2}\mathbb{E}[Z_{i}(1)|X_{i}=x]/\partial x^{2}$,
the components $q^{\prime}\mu_{-}^{(2)}$ and $q^{\prime}\mu_{+}^{(2)}$
in $\mathcal{B}$ become $\mu_{Y-}^{(2)}=\left.\partial^{2}\mathbb{E}[Y_{i}(0)|X_{i}=x]/\partial x^{2}\right|_{x=0}$
and $\mu_{Y+}^{(2)}=\left.\partial^{2}\mathbb{E}[Y_{i}(1)|X_{i}=x]/\partial x^{2}\right|_{x=0}$,
respectively. Thus, in this case, the convergence rates of the conventional
local polynomial estimators for $\mu_{Y-}^{(2)}$ and $\mu_{Y+}^{(2)}$
guarantee $\sqrt{\frac{nh_{n}^{5}}{\mathcal{V}}}(\bar{\mathcal{B}}-\mathcal{B})\overset{p}{\to}0$
(see, e.g., Fan and Gijbels, 1992, and Ruppert and Wand, 1994).} See Remark \ref{rem:V} below for a specific example of the variance
estimator $\bar{\mathcal{V}}$.

\begin{rem} {[}Efficiency comparison{]} It should be noted that the
asymptotic variance $\mathcal{V}$ in (\ref{eq:V}) of the estimator
$\bar{\tau}$ (or the bias corrected version $\bar{\tau}-h_{n}^{2}\bar{\mathcal{B}}$)
takes the same form as the one in CCFT even when the dimension of
$Z_{S^{*}}$ grows as $n$ increases. As investigated in CCFT, we
can see that the relative efficiency of $\bar{\tau}$ compared to
the conventional RDD estimator without covariates (say, $\hat{\tau}_{\mathrm{unadjusted}}$)
is
\begin{equation}
\frac{\mathcal{V}}{\mathcal{\mathcal{V}}_{\mathrm{unadjusted}}}=\frac{\sum_{t=0}^{1}\mathrm{Var}(Y_{i}(t)-Z_{i}(t)^{\prime}\gamma_{Y}|X_{i}=0)}{\sum_{t=0}^{1}\mathrm{Var}(Y_{i}(t)|X_{i}=0)},\label{eq:rel}
\end{equation}
where $\mathcal{\mathcal{V}}_{\mathrm{unadjusted}}$ is the asymptotic
variance of $\hat{\tau}_{\mathrm{unadjusted}}$ and $\gamma_{Y}$
is defined in (\ref{eq:gamY}). Generally there is no clear ranking
for these asymptotic variances. However, letting $\gamma_{Y,S^{*}}$
be an $s^{*}$-dimensional subvector of $\gamma_{Y}$ selected by
$S^{*}$, in an important special case where
\begin{equation}
\gamma_{Y,S^{*}}=\mathrm{Var}(Z_{S^{*},i}(t)|X_{i}=0)^{-1}\mathbb{E}[(Z_{S^{*},i}(t)-\mathbb{E}[Z_{S^{*},i}(t)|X_{i}])Y_{i}(t)|X_{i}=0]\quad\text{for }t=0\text{ and }1,\label{eq:gam}
\end{equation}
the coefficient vector $\gamma_{Y,S^{*}}$ becomes the best linear
approximation by $Z_{S^{*}}(t)$ for each group and thus $\bar{\tau}$
is asymptotically more efficient than $\hat{\tau}_{\mathrm{unadjusted}}$.

The relative efficiency in (\ref{eq:rel}) is also insightful to illustrate
the merit of our Lasso approach compared to CCFT. Let $Z=[Z_{S^{*}}:Z_{S^{*c}}]$
and $Z_{C}$ be covariates employed to apply the CCFT estimator $\hat{\tau}_{\mathrm{CCFT}}$
(without the Lasso covariates selection). Consider the special case
in (\ref{eq:gam}). As far as $Z_{C}$ contains $Z_{S^{*}}$, $\bar{\tau}$
and $\hat{\tau}_{\mathrm{CCFT}}$ achieve the same asymptotic efficiency
$\mathcal{V}$. On the other hand, if $Z_{C}$ does not contain some
elements of $Z_{S^{*}}$, then under (\ref{eq:gam}), the CCFT estimator
$\hat{\tau}_{\mathrm{CCFT}}$ is asymptotically less efficient than
the post-Lasso estimator $\bar{\tau}$. Therefore, when researchers
are less certain whether all elements of $Z_{S^{*}}$ are included
in $Z_{C}$ typically due to too many candidates in $Z$ or too small
effective sample sizes used for estimation, our Lasso-based approach
may be more attractive to achieve asymptotic efficiency $\mathcal{V}$
in more broader situations. We emphasize that such an efficiency gain
of our estimator $\bar{\tau}$ can be achieved at the cost of the
additional sparsity condition in Assumption \ref{asm:point-est},
which is trivially satisfied when the set of active covariates is
unknown but fixed. \end{rem}

\begin{rem} {[}Finite sample comparison{]} Although there is no gain
of using $\bar{\tau}$ instead of $\hat{\tau}_{\mathrm{CCFT}}$ in
terms of asymptotic efficiency as far as $Z_{C}$ contains $Z_{S^{*}}$,
the estimator $\bar{\tau}$ may exhibit better finite sample performance
even in such a scenario. To see this point, let $(\hat{\beta}_{C}^{\prime},\hat{\gamma}_{C}^{\prime})$
be the OLS estimator for the regression of $K^{1/2}Y$ on $K^{1/2}(1,T,X,TX)$
and $K^{1/2}Z_{C}$ so that $\hat{\tau}_{\mathrm{CCFT}}$ is the second
element of $\hat{\beta}_{C}$. As can be seen from CCFT's supplement,
one of the remainder terms of $\sqrt{\frac{nh_{n}}{\mathcal{V}}}(\hat{\tau}_{\mathrm{CCFT}}-\tau)$
involves a linear combination of the estimation error $\hat{\gamma}_{C}-(\gamma^{*\prime},0^{\prime})^{\prime}$.
Since the $\ell_{2}$-convergence rate of $\hat{\gamma}_{C}-(\gamma^{*\prime},0^{\prime})^{\prime}$
is typically of order $\sqrt{\dim Z_{C}/n}$, this remainder term
is of larger order than $\hat{\gamma}^{*}-\gamma_{Y}$, where $(\hat{\beta}^{*\prime},\hat{\gamma}^{*\prime})$
is the OLS estimator for the regression of $K^{1/2}Y$ on $K^{1/2}(1,T,X,TX)$
and $K^{1/2}Z_{S^{*}}$. Even though these remainder terms do not
appear in the first order asymptotic distribution, they contribute
to the finite sample behaviors of $\bar{\tau}$ and $\hat{\tau}_{\mathrm{CCFT}}$.
Another finite sample issue we encounter in our simulation study below
is that as the dimension of $Z_{C}$ increases, the values of the
MSE-optimal bandwidth tend to be smaller (due to larger bias estimates
but relatively stable variance estimates). Thus the effective sample
size used for the RDD estimation tends to be smaller so that we observe
larger variations in the resulting RDD estimates across simulation
draws. Finally, our simulation results in Section \ref{sec:sim} (particularly
DGPs 2 and 3 with large $p$) illustrate that the covariate selection
approach exhibits smaller standard deviations than CCFT in finite
samples even when the asymptotic variances may be equivalent. \end{rem}

\begin{rem} {[}Selection consistency{]} Although Theorem \ref{thm:t}
is our main result on inference of the causal effect $\tau$, it is
also possible to derive the consistency of the selection procedure
(i.e., $\mathbb{P}\{\hat{S}=S^{*}\}\to1$) under some additional $\beta$-min
type condition (i.e., there exists some $\varepsilon>0$ such that
$|\gamma_{j}^{*}|>\lambda_{n}\varrho_{n}s^{*}(1+\varepsilon)$ for
each $j\in S^{*}$). See a working paper version of this paper (Arai,
Otsu and Seo, 2021) for more details on the selection consistency.
\end{rem}

\begin{rem}\label{rem:V} {[}Estimation of $\mathcal{V}${]} An example
of the estimator $\bar{\mathcal{V}}$ for the asymptotic variance
$\mathcal{V}$ is the nearest neighborhood estimator, which is employed
in CCT, CCFT, and our numerical illustrations. Let
\begin{eqnarray*}
\bar{\varepsilon}_{V-,i} & = & \mathbb{I}\{X_{i}<0\}\sqrt{\frac{J}{J+1}}\left(V_{i}-\frac{1}{J}\sum_{j=1}^{J}V_{\ell_{-,j}(i)}\right),\\
\bar{\varepsilon}_{V+,i} & = & \mathbb{I}\{X_{i}\ge0\}\sqrt{\frac{J}{J+1}}\left(V_{i}-\frac{1}{J}\sum_{j=1}^{J}V_{\ell_{+,j}(i)}\right),
\end{eqnarray*}
for $V\in\{Y,Z_{1},\ldots,Z_{p}\}$ and a fixed positive integer $J$,
where $\ell_{-,j}(i)$ is the index of the $j$-th closest unit to
unit $i$ among $\{i:X_{i}<0\}$, and $\ell_{+,j}(i)$ is the index
of the $j$-th closest unit to unit $i$ among $\{i:X_{i}\ge0\}$.
The nearest neighborhood estimators of $\Sigma_{-}$ and $\Sigma_{+}$
in (\ref{eq:Vnote}) are defined as
\[
\bar{\Sigma}_{-}^{NN}=\left[\begin{array}{cccc}
\bar{\Sigma}_{YY-}^{NN} & \bar{\Sigma}_{YZ_{1}-}^{NN} & \cdots & \bar{\Sigma}_{YZ_{|\hat{S}|}-}^{NN}\\
\bar{\Sigma}_{Z_{1}Y-}^{NN} & \bar{\Sigma}_{Z_{1}Z_{1}-}^{NN}\\
\vdots &  & \ddots\\
\bar{\Sigma}_{Z_{|\hat{S}|}Y-}^{NN} &  &  & \bar{\Sigma}_{Z_{|\hat{S}|}Z_{|\hat{S}|}-}^{NN}
\end{array}\right],\quad\bar{\Sigma}_{+}^{NN}=\left[\begin{array}{cccc}
\bar{\Sigma}_{YY+}^{NN} & \bar{\Sigma}_{YZ_{1}+}^{NN} & \cdots & \bar{\Sigma}_{YZ_{|\hat{S}|}+}^{NN}\\
\bar{\Sigma}_{Z_{1}Y+}^{NN} & \bar{\Sigma}_{Z_{1}Z_{1}+}^{NN}\\
\vdots &  & \ddots\\
\bar{\Sigma}_{Z_{|\hat{S}|}Y+}^{NN} &  &  & \bar{\Sigma}_{Z_{|\hat{S}|}Z_{|\hat{S}|}+}^{NN}
\end{array}\right],
\]
where $\bar{\Sigma}_{VW-}^{NN}$ and $\bar{\Sigma}_{VW+}^{NN}$ are
$n\times n$ matrices whose $(i,j)$-th elements are
\begin{eqnarray*}
[\bar{\Sigma}_{VW-}^{NN}]_{i,j} & = & \mathbb{I}\{X_{i}<0\}\mathbb{I}\{X_{j}<0\}\mathbb{I}\{i=j\}\bar{\varepsilon}_{V-,i}\bar{\varepsilon}_{W-,j},\\{}
[\bar{\Sigma}_{VW+}^{NN}]_{i,j} & = & \mathbb{I}\{X_{i}\ge0\}\mathbb{I}\{X_{j}\ge0\}\mathbb{I}\{i=j\}\bar{\varepsilon}_{V+,i}\bar{\varepsilon}_{W+,j},
\end{eqnarray*}
for $i,j=1,\ldots,n$ and $V,W\in\{Y,Z_{1},\ldots,Z_{p}\}$. Then
nearest neighborhood estimator of $\mathcal{V}$ is defined as
\[
\bar{\mathcal{V}}^{NN}=(\bar{q}\otimes P_{-}^{\prime}e_{1})^{\prime}\bar{\Sigma}_{-}^{NN}(\bar{q}\otimes P_{-}^{\prime}e_{1})+(\bar{q}\otimes P_{+}^{\prime}e_{1})^{\prime}\bar{\Sigma}_{+}^{NN}(\bar{q}\otimes P_{+}^{\prime}e_{1}).
\]
By adapting the proof of Section S.2.4 in the supplement of CCT, the
consistency of this variance estimator is obtained as follows.

\begin{prop}\label{prop:V} Suppose that the assumptions of Theorem
\ref{thm:t} hold true. Additionally assume that $\mathrm{Var}((Y_{i}(t),Z_{i}(t)^{\prime})|X_{i}=x)$
is Lipschitz continuous in a neighborhood of $0$, and
\begin{align*}
 & \varpi_{n}^{-1}\{(m_{n}|S_{n}|\lambda_{n})\vee(|\hat{S}_{1}|\zeta_{n})\vee(m_{1n}|\hat{S}_{1}|\zeta_{n}^{2}/\lambda_{n})\}\overset{p}{\to}0,\\
 & (\varpi_{n}nh_{n})^{-1}s^{*}\sqrt{\log s^{*}}\overset{p}{\to}0,\qquad\zeta_{n}^{-1}\frac{\lambda_{n}s^{*}}{\phi^{*2}}\to0,
\end{align*}
where $\varpi_{n}$ is defined in (\ref{pf:omega}). Then $\frac{\bar{\mathcal{V}}^{NN}}{\mathcal{V}}\overset{p}{\to}1$.
\end{prop} \end{rem}

\section{Discussion\label{sec:dis}}

\subsection{Fuzzy RDD\label{sub:fuzzy}}

Although the discussion so far focuses on the sharp RDD analysis,
it is possible to extend our approach to the fuzzy RDD analysis, where
the forcing variable $X_{i}$ is not informative enough to determine
the treatment $W_{i}$ but still affects the treatment probability.
In particular, the fuzzy RDD assumes that the conditional treatment
probability $\mathbb{P}\{W_{i}=1|X_{i}=x\}$ jumps at the cutoff point
$\bar{x}$. As in the last section, we normalize $\bar{x}=0$. To
define a reasonable parameter of interest for the fuzzy case, let
$W_{i}(x)$ be a potential treatment for unit $i$ when the cutoff
level for the treatment is set at $x$, and assume that $W_{i}(x)$
is non-increasing in $x$ at $x=0$. Using the terminology of Angrist,
Imbens and Rubin (1996), unit $i$ is called a complier if her cutoff
level is $X_{i}$ (i.e., $\lim_{x\downarrow X_{i}}W_{i}(x)=0$ and
$\lim_{x\uparrow X_{i}}W_{i}(x)=1$). A parameter of interest in the
fuzzy RDD, suggested by Hahn, Todd and van der Klaauw (2001), is the
average causal effect for compliers at $x=0$, 
\[
\tau_{f}=\mathbb{E}[Y_{i}(1)-Y_{i}(0)|i\mbox{ is complier, }X_{i}=0].
\]
 Hahn, Todd and van der Klaauw (2001) showed that under mild conditions
the parameter $\tau_{f}$ can be identified by the ratio of the jump
in the conditional mean of $Y_{i}$ at $x=0$ to the jump in the conditional
treatment probability at $X_{i}=0$, i.e.,
\begin{equation}
\tau_{f}=\frac{\lim_{x\downarrow0}\mathbb{E}[Y_{i}|X_{i}=x]-\lim_{x\uparrow0}\mathbb{E}[Y_{i}|X_{i}=x]}{\lim_{x\downarrow0}\mathbb{P}\{W_{i}=1|X_{i}=x\}-\lim_{x\uparrow0}\mathbb{P}\{W_{i}=1|X_{i}=x\}}.\label{eq:fuzzy}
\end{equation}

In this case, letting $T_{i}=\mathbb{I}\{X_{i}\ge0\}$, the numerator
and denominator of (\ref{eq:fuzzy}) can be estimated by the local
Lasso estimators $\hat{\vartheta}_{1}=(\hat{\alpha}_{1},\hat{\tau}_{1},\hat{\beta}_{1-},\hat{\beta}_{1+},\hat{\gamma}_{1}^{\prime})^{\prime}$
and $\hat{\vartheta}_{2}=(\hat{\alpha}_{2},\hat{\tau}_{2},\hat{\beta}_{2-},\hat{\beta}_{2+},\hat{\gamma}_{2}^{\prime})^{\prime}$,
which solve
\begin{align*}
 & \min_{\alpha_{1},\tau_{1},\beta_{1-},\beta_{1+},\gamma_{1}}\frac{1}{nb_{1n}}\sum_{i=1}^{n}K\left(\frac{X_{i}}{b_{1n}}\right)(Y_{i}-\alpha_{1}-T_{i}\tau_{1}-X_{i}\beta_{1-}-T_{i}X_{i}\beta_{1+}-Z_{1i}^{\prime}\gamma_{1})^{2}+\lambda_{1n}|\gamma_{1}|_{1},\\
 & \min_{\alpha_{2},\tau_{2},\beta_{2-},\beta_{2+},\gamma_{2}}\frac{1}{nb_{2n}}\sum_{i=1}^{n}K\left(\frac{X_{i}}{b_{2n}}\right)(W_{i}-\alpha_{2}-T_{i}\tau_{2}-X_{i}\beta_{2-}-T_{i}X_{i}\beta_{2+}-Z_{2i}^{\prime}\gamma_{2})^{2}+\lambda_{2n}|\gamma_{2}|_{1},
\end{align*}
where $Z_{1i}$ and $Z_{2i}$ are predetermined covariates used for
the first and second regressions, respectively. Then, based on the
selected covariates from the above local Lasso regressions (i.e.,
$\tilde{S}_{1}=\{j:|\hat{\gamma}_{1j}|>\zeta_{n}\}$ with a subvector
$Z_{1\tilde{S},i}$ of $Z_{1i}$ selected by $\tilde{S}_{1}$, and
$\tilde{S}_{2}=\{j:|\hat{\gamma}_{2j}|>\zeta_{n}\}$ with a subvector
$Z_{2\tilde{S},i}$ of $Z_{2i}$ selected by $\tilde{S}_{2}$), we
implement the local least squares as in CCFT:
\begin{align*}
 & \min_{\alpha_{1},\tau_{1},\beta_{1-},\beta_{1+},t_{1}}\frac{1}{nh_{1n}}\sum_{i=1}^{n}K\left(\frac{X_{i}}{h_{1n}}\right)(Y_{i}-\alpha_{1}-T_{i}\tau_{1}-X_{i}\beta_{1-}-T_{i}X_{i}\beta_{1+}-Z_{1\tilde{S},i}^{\prime}t_{1})^{2},\\
 & \min_{\alpha_{2},\tau_{2},\beta_{2-},\beta_{2+},t_{2}}\frac{1}{nh_{2n}}\sum_{i=1}^{n}K\left(\frac{X_{i}}{h_{2n}}\right)(W_{i}-\alpha_{2}-T_{i}\tau_{2}-X_{i}\beta_{2-}-T_{i}X_{i}\beta_{2+}-Z_{2\tilde{S},i}^{\prime}t_{2})^{2}.
\end{align*}
The numerator and denominator of (\ref{eq:fuzzy}) are given by the
estimated coefficients of $\tau_{1}$ and $\tau_{2}$ for the above
minimizations, respectively. If the treatment variable $W$ satisfies
analogous conditions for Theorem \ref{thm:point-est} (by replacing
$W$ with $Y$), we expect that analogous results to the sharp RDD
case can be established.

\subsection{Regression kink design\label{sub:RKD}}

Our high-dimensional method can be extended for the RKDs. For each
unit $i=1,\ldots,n$, we observe continuous outcome and explanatory
variables denoted by $Y_{i}$ and $X_{i}$, respectively. The RKD
analysis is concerned with the following nonseparable model 
\[
Y=f(Q,X,U),
\]
where $U$ is an error term (possibly multivariate) and $Q=q(X)$
is a continuous policy variable of interest with known $q(\cdot)$.
In general, even though we know the function $q(\cdot)$, we are not
able to identify the treatment effect by the policy variable $Q$.
However, it is often the case that the policy function $q(\cdot)$
has some kinks (but is continuous). For instance, suppose $Y$ is
duration of unemployment and $X$ is earnings before losing the job.
We are interested in the effect of unemployment benefits $Q=q(X)$.
In many unemployment insurance systems (e.g., the one in Austria),
$q(\cdot)$ is specified by a piecewise linear function. In such a
scenario, one may exploit changes of slopes in the conditional mean
$\mathbb{E}[Y|X=x]$ to identify a treatment effect of $Q$. Suppose
$q(\cdot)$ is kinked at $0$. Otherwise, we redefine $X$ by subtracting
the kink point $c$ from $X$. In particular, Card, \emph{et al.}
(2015) showed that a treatment on treated parameter $\tau_{k}=\int\frac{\partial f(q,x,u)}{\partial q}dF_{U|Q=q,X=x}(u)$
is identified as 
\begin{equation}
\tau_{k}=\frac{\lim_{x\downarrow0}\frac{d}{dx}\mathbb{E}[Y|X=x]-\lim_{x\uparrow0}\frac{d}{dx}\mathbb{E}[Y|X=x]}{\lim_{x\downarrow0}\frac{d}{dx}q(x)-\lim_{x\uparrow0}\frac{d}{dx}q(x)}.\label{eq:tau0}
\end{equation}
To estimate $\tau_{k}$, we propose the following local Lasso regression
\begin{equation}
\min_{\alpha,\delta,\beta,\zeta,\eta,\gamma}\frac{1}{nb_{n}}\sum_{i=1}^{n}K\left(\frac{X_{i}}{b_{n}}\right)(Y_{i}-\alpha-T_{i}X_{i}\delta-X_{i}\beta-T_{i}X_{i}^{2}\zeta-X_{i}^{2}\eta-Z_{i}^{\prime}\gamma)^{2}+\lambda_{n}|\gamma|_{1}.\label{eq:lasso-1}
\end{equation}
Let $\hat{\delta}$ be the Lasso estimator of $\delta$ by (\ref{eq:lasso-1}).
Since the denominator $q_{0}=\lim_{x\downarrow c}\frac{d}{dx}q(x)-\lim_{x\uparrow c}\frac{d}{dx}q(x)$
in (\ref{eq:tau0}) is assumed to be known, the estimator of $\tau_{k}$
is given by $\hat{\tau}_{k}=\hat{\delta}/q_{0}$. Under analogous
conditions to Theorem \ref{thm:point-est}, we expect that analogous
results to the sharp RDD case can be established.

\section{Recommendation for Implementation\label{sec:rec}}

We recommend the following steps to implement our method in R. The
most important difference from CCFT is the covariate selection step
(Step 2). This procedure is employed in our numerical studies in the
following sections.
\begin{enumerate}
\item Without using data on covariates, obtain the MSE-optimal bandwidth
$b_{n}=b_{n}^{\mathrm{CCT}}$ developed by CCT. This can be implemented
by an R package, \texttt{Rdrobust} (Calonico, Cattaneo and Titiunik,
2015b).
\item Using the data on covariates, implement the Lasso estimation in (\ref{eq:lasso})
to compute $\hat{\theta}$ by setting $b_{n}=b_{n}^{\mathrm{CCT}}$
and choosing $\lambda_{n}$ by a data-driven choice. This can be implemented
by an R package, \texttt{hdm} (https://cran.r-project.org/web/packages/hdm/index.html)
for example.
\item Based on the Lasso estimates $\hat{\theta}$ obtained in Step 2, select
the covariates $Z_{\hat{S},i}$ by using the trimming term $\zeta_{n}$
in (\ref{eq:zeta}). Then implement CCFT's RDD estimation by (\ref{eq:post-lasso})
and inference procedure including the bias correction and bandwidth
selection. For this step, we recommend to follow the procedures detailed
in CCFT's supplement by using \texttt{Rdrobust}.
\end{enumerate}
\noindent The MSE-optimal bandwidth with the full set of covariates
can be too narrow when the number of covariates is large relative
to the number of observations and many of covariates are not relevant.
Given that we do not know the exact identities of $S^{*}$ a priori,
we instead employ the MSE-optimal bandwidth by CCT in Step 1, which
does not include any covariate. This yields more robust results as
shown in the simulation in the next section. Furthermore, we note
that CCFT's bandwidth selection in Step 3 is derived under the assumption
of fixed dimensional covariates even though the dimension of the selected
covariates $Z_{\hat{S},i}$ may grow. We recommend the above procedure
because of its practicality given the available packages. Further
theoretical analysis for an optimal choice of the bandwidth parameters
and other tuning constants in the present setup is beyond the scope
of this paper. We note that the choice described in the above procedure
works well for all data generating processes considered in our numerical
studies.

\section{Simulation\label{sec:sim}}

In this section, we conduct simulation experiments to investigate
finite sample properties of our covariate selection approach for estimation
and inference on the sharp RDD analysis. We consider three simulation
designs based on CCFT with introducing additional covariates. The
additional covariates are generated based on the simulation designs
in Belloni, Chernozhukov and Hansen (2014). Let $\mathcal{B}(a,b)$
be a beta distribution with parameters $a$ and $b$. The data generating
process (DGP) is specified as follows
\[
Y=\mu_{1}(X)+\mu_{2}(Z)+\mu_{3}(W)+\varepsilon_{y},\quad X\sim2\mathcal{B}(2,4)-1,\quad Z=\mu_{z}(X)+\varepsilon_{z},
\]
\[
\mu_{z}(x)=\begin{cases}
0.49+1.06x+5.74x^{2}+17.14x^{3}+19.75x^{5}+7.47x^{5} & \mbox{for }x<0,\\
0.49+0.61x+0.23x^{2}-3.46x^{3}+6.43x^{4}-3.48x^{5} & \mbox{for }x\ge0,
\end{cases}
\]
$W=(W_{1},\ldots,W_{p})^{\prime}\sim N(0,\Sigma_{W})$ with $\mathbb{E}(W_{l}^{2})=1$
and $\mathrm{Cov}(W_{l,}W_{l_{1}})=0.5^{|l-l_{1}|}$, and

\[
\left(\begin{array}{c}
\varepsilon_{y}\\
\varepsilon_{z}
\end{array}\right)\sim N(0,\Sigma),\quad\Sigma=\left(\begin{array}{cc}
\sigma_{y}^{2} & \rho\sigma_{y}\sigma_{z}\\
\rho\sigma_{y}\sigma_{z} & \sigma_{z}^{2}
\end{array}\right),
\]
with $\sigma_{y}=0.1295$ and $\sigma_{z}=0.1353$.

For the functions $\mu_{1}$, $\mu_{2}$, and $\mu_{3}$, we consider
three cases. For DGP1, we set $\rho=0.2692$, all the coefficients
of $\mu_{2}(z)$ and $\mu_{3}(w)$ to be zero, and
\[
\mu_{1}(x)=\begin{cases}
0.48+1.27x+7.18x^{2}+20.21x^{3}+21.54x^{4}+7.33x^{5} & \mbox{for }x<0,\\
0.52+0.84x-3.00x^{2}+7.99x^{3}-9.01x^{4}+3.56x^{5} & \mbox{for }x\ge0,
\end{cases}
\]
For DGP2, we set $\rho=0.2692$,
\[
\mu_{1}(x)=\begin{cases}
0.36+0.96x+5.47x^{2}+15.28x^{3}+15.87x^{4}+5.14x^{5} & \mbox{for }x<0,\\
0.38+0.62x-2.84x^{2}+8.42x^{3}-10.24x^{4}+4.31x^{5} & \mbox{for }x\ge0,
\end{cases}
\]

\[
\mu_{2}(z)=\begin{cases}
0.22z & \mbox{for }x<0,\\
0.28z & \mbox{for }x\ge0,
\end{cases}
\]
and $\mu_{3}(w)=\sum_{l=0}^{p-1}\pi_{l}w_{l}$ with $\pi_{l}=0.2^{l}$.
For DGP3, we set $\mu_{1}(x)$ and $\mu_{2}(z)$ as in DGP2, and $\mu_{3}(w)=\sum_{l=0}^{p-1}\pi_{l}w_{l}$
with $\pi_{l}=0.5^{l}$.\footnote{Strictly speaking, DGPs 2 and 3 do not satisfy the sparsity assumption
(when $p$ is large) because the coefficients $\pi_{l}$ are not exactly
equal to zero even for very large\textbf{ $l$}. However, our unreported
simulation results show that the variable selection approach produces
identical results to Tables 1-4 when the DGPs are modified to satisfy
the exact sparsity assumption, where $\pi_{l}$ are set to zero for
$l>n^{9/10}$. Because the covariate adjusted approaches are unstable,
their results differ but are similar qualitatively.} DGP1 is the baseline model considered by CCFT which contains no covariates.
DGPs 2 and 3 introduce additional covariates to Model 2 of CCFT with
different degrees of magnitude in the coefficients of $z$. For DGP4,
we set $\mu_{1}(x)$ and $\mu_{2}(z)$ as in DGP2, and $\mu_{3}(w)=\sum_{l=1}^{p}\pi_{l}w_{l}$
with $\pi_{l}=0.2l^{-1}$. DGP4 is considered to see how all the approaches
behave when the (approximate) sparsity assumption fails. The sample
size is set as $n=500$ for all cases. The number of the covariates
$p$ varies from 5 to 500. The results are based on 1,000 Monte Carlo
replications.

Table \ref{tab:point-est} shows the biases and RMSEs of the following
four point estimation methods. For the bandwidth $h_{n}$, the first
two methods use the MSE-optimal bandwidth without covariates proposed
by CCT. The third method uses the MSE-optimal bandwidth with covariates
proposed by CCFT. The fourth method, called ``Adaptive'', is the
bandwidth for our covariate selection approach which uses the MSE-optimal
bandwidth without covariates for the covariate selection stage and
uses that with covariates in the estimation stage. For estimation
methods, the first method uses the standard RDD estimation method
without covariates by CCT. The second and the third methods use the
RDD estimation with covariates by CCFT. The fourth method uses the
RDD estimation with the selected covariates. See Section \ref{sec:rec}
for more details on the fourth method.

Our findings are summarized as follows. First, the RMSEs of the covariate
adjusted estimation get larger irrespective of the bandwidths as the
number of covariates increases across all DPGs. These increases in
the RMSEs are due to inflated standard errors caused by a large number
of covariates. This result clearly indicates the need for covariate
selection. Second, the covariate selection approach shows excellent
performances for all cases. Both the biases and RMSEs are stable for
different values of $p$ for all designs. Finally, all methods work
equally well for DGP1, where all the additional covariates are irrelevant.
However, for DGP2 and DGP3, we find substantial efficiency loss of
the standard method. The results for DGP4 show that the standard and
covariate selection approaches preform stably even when the sparsity
assumption is not satisfied. However, the covariate adjusted approaches
by CCFT perform unstably in terms of the RMSE. Overall, we recommend
the covariate selection even for relatively small $p$.

Table \ref{tab:selected} reports the number of selected covariates
for our covariate selection approach. It is quite natural that the
number of selected covariates increases when the number of non-zero
coefficients of covariates increases. It is interesting to note that
the average number of selected covariates decreases as the number
of covariates increase.

Table \ref{tab:inf} shows the coverage probabilities and interval
lengths of the robust confidence intervals for the causal effect.
The nominal coverage level is $0.95$. The following points are notable.
First, the performances of our covariate selection approach are stable
for all DGPs, although the coverage probabilities tend to be a little
bit smaller than the nominal level. Second, for the covariate adjusted
approaches, the coverage probabilities decrease and the interval lengths
get shorter as $p$ increases. Third, the coverage of CCT is more
stable and better than other methods, especially for DGP2 and DGP3.
However, the average lengths of CCT are substantially longer than
the other methods. Fourth, even when the sparsity assumption fails
as in DGP4, the standard and covariate selection approaches preform
stably but the covariate adjusted approach exhibits erratic behaviors
in terms of the coverages and interval lengths. Overall, the covariate
selection approach is promising for inference as well since it exhibits
robust performances in both coverages and lengths for different numbers
of covariates and DGPs.

Finally, Table \ref{tab:band} presents the properties of the MSE-optimal
bandwidths. We observe that the MSE-optimal bandwidth without covariates
and the adaptive one are very stable while the MSE-optimal bandwidth
with covariates shrinks as $p$ increases. This is possibly the main
source of the increased RMSE and under-coverages.

We note that the unstable performance of the covariate adjusted approaches
when the number of covariates is large can be because their optimal
bandwidths are not available for cases where the number of covariates
is large. It would be more appropriate to develop optimal bandwidths
for the case of high-dimensional covariates and make comparisons,
but it is beyond the scope of this paper.

\newpage{}

\begin{table}[htp]
{\small{}\caption{Simulation: Point estimation{\small{}\label{tab:point-est}}}
}{\small\par}
\centering{}{\small{}}%
\begin{tabular}{lccccccccc}
\hline 
\multicolumn{2}{c}{{\small{}MSE-optimal bandwidths:}} & \multicolumn{4}{c}{{\small{}w/o Covariates}} & \multicolumn{2}{c}{{\small{}w/ Covariates}} & \multicolumn{2}{c}{{\small{}Adaptive}}\tabularnewline
\multicolumn{2}{c}{{\small{}Estimation methods:}} & \multicolumn{2}{c}{{\small{}Standard}} & \multicolumn{2}{c}{{\small{}Covariate adjusted}} & \multicolumn{2}{c}{{\small{}Covariate adjusted}} & \multicolumn{2}{c}{{\small{}Covariate selection}}\tabularnewline
 & {\small{}$p$} & {\small{}Bias} & {\small{}RMSE} & {\small{}Bias} & {\small{}RMSE} & {\small{}Bias} & {\small{}RMSE} & {\small{}Bias} & {\small{}RMSE}\tabularnewline
\hline 
{\small{}DGP1} & {\small{}5} & {\small{}0.022} & {\small{}0.062} & {\small{}0.023} & {\small{}0.064} & {\small{}0.022} & {\small{}0.065} & {\small{}0.022} & {\small{}0.062}\tabularnewline
 & {\small{}10} & {\small{}0.021} & {\small{}0.065} & {\small{}0.022} & {\small{}0.068} & {\small{}0.019} & {\small{}0.070} & {\small{}0.020} & {\small{}0.065}\tabularnewline
 & {\small{}20} & {\small{}0.020} & {\small{}0.061} & {\small{}0.019} & {\small{}0.066} & {\small{}0.013} & {\small{}0.074} & {\small{}0.019} & {\small{}0.061}\tabularnewline
 & {\small{}30} & {\small{}0.021} & {\small{}0.064} & {\small{}0.022} & {\small{}0.074} & {\small{}0.018} & {\small{}0.089} & {\small{}0.020} & {\small{}0.064}\tabularnewline
 & {\small{}40} & {\small{}0.022} & {\small{}0.064} & {\small{}0.022} & {\small{}0.080} & {\small{}0.002} & {\small{}0.323} & {\small{}0.022} & {\small{}0.064}\tabularnewline
 & {\small{}50} & {\small{}0.021} & {\small{}0.065} & {\small{}0.027} & {\small{}0.101} & {\small{}0.003} & {\small{}0.316} & {\small{}0.021} & {\small{}0.065}\tabularnewline
 & {\small{}100} & {\small{}0.019} & {\small{}0.063} & {\small{}0.038} & {\small{}0.387} & {\small{}0.004} & {\small{}0.133} & {\small{}0.018} & {\small{}0.063}\tabularnewline
 & {\small{}250} & {\small{}0.018} & {\small{}0.065} & {\small{}0.026} & {\small{}0.082} & {\small{}0.006} & {\small{}0.102} & {\small{}0.017} & {\small{}0.065}\tabularnewline
 & {\small{}500} & {\small{}0.021} & {\small{}0.064} & {\small{}0.028} & {\small{}0.072} & {\small{}0.009} & {\small{}0.100} & {\small{}0.020} & {\small{}0.064}\tabularnewline
\hline 
{\small{}DGP2} & {\small{}5} & {\small{}0.019} & {\small{}0.540} & {\small{}0.000} & {\small{}0.058} & {\small{}0.029} & {\small{}0.081} & {\small{}0.029} & {\small{}0.090}\tabularnewline
 & {\small{}10} & {\small{}-0.007} & {\small{}0.525} & {\small{}-0.001} & {\small{}0.061} & {\small{}0.026} & {\small{}0.084} & {\small{}0.028} & {\small{}0.094}\tabularnewline
 & {\small{}20} & {\small{}0.008} & {\small{}0.543} & {\small{}-0.007} & {\small{}0.064} & {\small{}0.020} & {\small{}0.094} & {\small{}0.026} & {\small{}0.099}\tabularnewline
 & {\small{}30} & {\small{}-0.006} & {\small{}0.530} & {\small{}-0.006} & {\small{}0.070} & {\small{}0.025} & {\small{}0.123} & {\small{}0.027} & {\small{}0.105}\tabularnewline
 & {\small{}40} & {\small{}0.002} & {\small{}0.521} & {\small{}-0.012} & {\small{}0.080} & {\small{}-0.001} & {\small{}1.485} & {\small{}0.026} & {\small{}0.102}\tabularnewline
 & {\small{}50} & {\small{}0.043} & {\small{}0.531} & {\small{}-0.010} & {\small{}0.098} & {\small{}0.002} & {\small{}1.116} & {\small{}0.025} & {\small{}0.105}\tabularnewline
 & {\small{}100} & {\small{}0.042} & {\small{}0.544} & {\small{}-0.033} & {\small{}0.551} & {\small{}-0.052} & {\small{}0.874} & {\small{}0.023} & {\small{}0.107}\tabularnewline
 & {\small{}250} & {\small{}-0.007} & {\small{}0.505} & {\small{}-0.059} & {\small{}1.162} & {\small{}0.027} & {\small{}1.166} & {\small{}0.019} & {\small{}0.116}\tabularnewline
 & {\small{}500} & {\small{}0.003} & {\small{}0.535} & {\small{}-0.058} & {\small{}1.636} & {\small{}-0.066} & {\small{}1.578} & {\small{}0.027} & {\small{}0.118}\tabularnewline
\hline 
{\small{}DGP3} & {\small{}5} & {\small{}0.013} & {\small{}0.708} & {\small{}-0.004} & {\small{}0.058} & {\small{}0.029} & {\small{}0.081} & {\small{}0.028} & {\small{}0.103}\tabularnewline
 & {\small{}10} & {\small{}-0.010} & {\small{}0.681} & {\small{}-0.005} & {\small{}0.060} & {\small{}0.026} & {\small{}0.084} & {\small{}0.030} & {\small{}0.179}\tabularnewline
 & {\small{}20} & {\small{}0.002} & {\small{}0.706} & {\small{}-0.011} & {\small{}0.064} & {\small{}0.020} & {\small{}0.094} & {\small{}0.005} & {\small{}0.177}\tabularnewline
 & {\small{}30} & {\small{}-0.009} & {\small{}0.691} & {\small{}-0.011} & {\small{}0.070} & {\small{}0.025} & {\small{}0.123} & {\small{}0.030} & {\small{}0.186}\tabularnewline
 & {\small{}40} & {\small{}0.000} & {\small{}0.683} & {\small{}-0.017} & {\small{}0.080} & {\small{}0.021} & {\small{}0.321} & {\small{}0.018} & {\small{}0.176}\tabularnewline
 & {\small{}50} & {\small{}0.038} & {\small{}0.686} & {\small{}-0.015} & {\small{}0.085} & {\small{}0.042} & {\small{}0.784} & {\small{}0.017} & {\small{}0.190}\tabularnewline
 & {\small{}100} & {\small{}0.048} & {\small{}0.722} & {\small{}-0.036} & {\small{}0.301} & {\small{}0.045} & {\small{}0.605} & {\small{}0.013} & {\small{}0.201}\tabularnewline
 & {\small{}250} & {\small{}-0.019} & {\small{}0.668} & {\small{}-0.055} & {\small{}0.347} & {\small{}-0.021} & {\small{}0.957} & {\small{}0.009} & {\small{}0.216}\tabularnewline
 & {\small{}500} & {\small{}-0.004} & {\small{}0.707} & {\small{}-0.050} & {\small{}0.485} & {\small{}0.015} & {\small{}0.965} & {\small{}0.014} & 0.217\tabularnewline
\hline 
{\small{}DGP4} & {\small{}5} & {\small{}0.028} & {\small{}0.180} & {\small{}0.018} & {\small{}0.066} & {\small{}0.029} & {\small{}0.081} & {\small{}0.022} & {\small{}0.131}\tabularnewline
 & {\small{}10} & {\small{}0.021} & {\small{}0.187} & {\small{}0.015} & {\small{}0.068} & {\small{}0.026} & {\small{}0.084} & {\small{}0.019} & {\small{}0.185}\tabularnewline
 & {\small{}20} & {\small{}0.019} & {\small{}0.193} & {\small{}0.010} & {\small{}0.071} & {\small{}0.020} & {\small{}0.094} & {\small{}0.015} & {\small{}0.192}\tabularnewline
 & {\small{}30} & {\small{}0.028} & {\small{}0.197} & {\small{}0.013} & {\small{}0.080} & {\small{}0.025} & {\small{}0.123} & {\small{}0.024} & {\small{}0.196}\tabularnewline
 & {\small{}40} & {\small{}0.021} & {\small{}0.192} & {\small{}0.008} & {\small{}0.092} & {\small{}0.001} & {\small{}1.480} & {\small{}0.019} & {\small{}0.190}\tabularnewline
 & {\small{}50} & {\small{}0.029} & {\small{}0.198} & {\small{}0.011} & {\small{}0.110} & {\small{}-0.026} & {\small{}1.491} & {\small{}0.022} & {\small{}0.194}\tabularnewline
 & {\small{}100} & {\small{}0.036} & {\small{}0.206} & {\small{}-0.053} & {\small{}0.853} & {\small{}-0.024} & {\small{}0.470} & {\small{}0.032} & {\small{}0.203}\tabularnewline
 & {\small{}250} & {\small{}0.017} & {\small{}0.194} & {\small{}-0.039} & {\small{}0.547} & {\small{}-0.072} & {\small{}0.578} & {\small{}0.013} & {\small{}0.190}\tabularnewline
 & {\small{}500} & {\small{}0.020} & {\small{}0.196} & {\small{}-0.026} & {\small{}0.549} & {\small{}-0.053} & {\small{}0.627} & {\small{}0.016} & {\small{}0.194}\tabularnewline
\hline 
\end{tabular}{\small\par}
\end{table}

\newpage{}

\begin{table}[htp]
{\small{}\caption{{\small{}\label{tab:selected}}Simulation: Number of selected covariates}
}{\small\par}
\centering{}{\small{}}%
\begin{tabular}{lcccc}
\hline 
 & {\small{}$p$} & {\small{}Average} & {\small{}Min} & {\small{}Max}\tabularnewline
\hline 
{\small{}DGP1} & {\small{}5} & {\small{}0.376} & {\small{}0} & {\small{}1}\tabularnewline
 & {\small{}10} & {\small{}0.362} & {\small{}0} & {\small{}1}\tabularnewline
 & {\small{}20} & {\small{}0.357} & {\small{}0} & {\small{}1}\tabularnewline
 & {\small{}30} & {\small{}0.353} & {\small{}0} & {\small{}1}\tabularnewline
 & {\small{}40} & {\small{}0.332} & {\small{}0} & {\small{}1}\tabularnewline
 & {\small{}50} & {\small{}0.379} & {\small{}0} & {\small{}1}\tabularnewline
 & {\small{}100} & {\small{}0.350} & {\small{}0} & {\small{}1}\tabularnewline
 & {\small{}250} & {\small{}0.331} & {\small{}0} & {\small{}1}\tabularnewline
 & {\small{}500} & {\small{}0.264} & {\small{}0} & {\small{}1}\tabularnewline
\hline 
{\small{}DGP2} & {\small{}5} & {\small{}2.868} & {\small{}2} & {\small{}3}\tabularnewline
 & {\small{}10} & {\small{}2.780} & {\small{}2} & {\small{}3}\tabularnewline
 & {\small{}20} & {\small{}2.689} & {\small{}1} & {\small{}3}\tabularnewline
 & {\small{}30} & {\small{}2.631} & {\small{}2} & {\small{}3}\tabularnewline
 & {\small{}40} & {\small{}2.555} & {\small{}1} & {\small{}3}\tabularnewline
 & {\small{}50} & {\small{}2.574} & {\small{}1} & {\small{}3}\tabularnewline
 & {\small{}100} & {\small{}2.478} & {\small{}1} & {\small{}3}\tabularnewline
 & {\small{}250} & {\small{}2.341} & {\small{}1} & {\small{}3}\tabularnewline
 & {\small{}500} & {\small{}2.303} & {\small{}1} & {\small{}3}\tabularnewline
\hline 
{\small{}DGP3} & {\small{}5} & {\small{}3.910} & {\small{}3} & {\small{}5}\tabularnewline
 & {\small{}10} & {\small{}3.063} & {\small{}2} & {\small{}5}\tabularnewline
 & {\small{}20} & {\small{}3.045} & {\small{}2} & {\small{}5}\tabularnewline
 & {\small{}30} & {\small{}3.003} & {\small{}1} & {\small{}4}\tabularnewline
 & {\small{}40} & {\small{}2.959} & {\small{}1} & {\small{}5}\tabularnewline
 & {\small{}50} & {\small{}2.920} & {\small{}1} & {\small{}5}\tabularnewline
 & {\small{}100} & {\small{}2.850} & {\small{}1} & {\small{}4}\tabularnewline
 & {\small{}250} & {\small{}2.683} & {\small{}1} & {\small{}4}\tabularnewline
 & {\small{}500} & {\small{}2.547} & {\small{}1} & {\small{}4}\tabularnewline
\hline 
{\small{}DGP4} & {\small{}5} & {\small{}1.771} & {\small{}1} & {\small{}3}\tabularnewline
 & {\small{}10} & {\small{}0.983} & {\small{}0} & {\small{}3}\tabularnewline
 & {\small{}20} & {\small{}0.860} & {\small{}0} & {\small{}3}\tabularnewline
 & {\small{}30} & {\small{}0.787} & {\small{}0} & {\small{}3}\tabularnewline
 & {\small{}40} & {\small{}0.719} & {\small{}0} & {\small{}3}\tabularnewline
 & {\small{}50} & {\small{}0.730} & {\small{}0} & {\small{}3}\tabularnewline
 & {\small{}100} & {\small{}0.680} & {\small{}0} & {\small{}3}\tabularnewline
 & {\small{}250} & {\small{}0.593} & {\small{}0} & {\small{}3}\tabularnewline
 & {\small{}500} & {\small{}0.479} & {\small{}0} & {\small{}3}\tabularnewline
\hline 
\end{tabular}{\small\par}
\end{table}

\newpage{}

\begin{table}[htp]
{\small{}\caption{{\small{}\label{tab:inf}}Simulation: Inference}
}{\small\par}
\centering{}{\small{}}%
\begin{tabular}{lccccccccc}
\hline 
\multicolumn{2}{c}{{\small{}MSE-optimal bandwidths:}} & \multicolumn{4}{c}{{\small{}w/o Covariates}} & \multicolumn{2}{c}{{\small{}w/ Covariates}} & \multicolumn{2}{c}{{\small{}Adaptive}}\tabularnewline
\multicolumn{2}{c}{{\small{}Estimation methods:}} & \multicolumn{2}{c}{{\small{}Standard}} & \multicolumn{2}{c}{{\small{}Covariate adjusted}} & \multicolumn{2}{c}{{\small{}Covariate adjusted}} & \multicolumn{2}{c}{{\small{}Covariate selection}}\tabularnewline
 & {\small{}$p$} & {\small{}CP} & {\small{}Length} & {\small{}CP} & {\small{}Length} & {\small{}CP} & {\small{}Length} & {\small{}CP} & {\small{}Length}\tabularnewline
\hline 
{\small{}DGP1} & {\small{}5} & {\small{}0.918} & {\small{}0.270} & {\small{}0.898} & {\small{}0.246} & {\small{}0.901} & {\small{}0.246} & {\small{}0.916} & {\small{}0.270}\tabularnewline
 & {\small{}10} & {\small{}0.892} & {\small{}0.213} & {\small{}0.855} & {\small{}0.211} & {\small{}0.856} & {\small{}0.205} & {\small{}0.891} & {\small{}0.213}\tabularnewline
 & {\small{}20} & {\small{}0.925} & {\small{}0.293} & {\small{}0.815} & {\small{}0.259} & {\small{}0.779} & {\small{}0.280} & {\small{}0.919} & {\small{}0.275}\tabularnewline
 & {\small{}30} & {\small{}0.914} & {\small{}0.260} & {\small{}0.734} & {\small{}0.148} & {\small{}0.658} & {\small{}0.156} & {\small{}0.914} & {\small{}0.260}\tabularnewline
 & {\small{}40} & {\small{}0.915} & {\small{}0.241} & {\small{}0.649} & {\small{}0.182} & {\small{}0.476} & {\small{}0.107} & {\small{}0.916} & {\small{}0.241}\tabularnewline
 & {\small{}50} & {\small{}0.915} & {\small{}0.246} & {\small{}0.540} & {\small{}0.152} & {\small{}0.247} & {\small{}0.076} & {\small{}0.917} & {\small{}0.246}\tabularnewline
 & {\small{}100} & {\small{}0.913} & {\small{}0.246} & {\small{}0.194} & {\small{}0.342} & {\small{}0.186} & {\small{}0.040} & {\small{}0.911} & {\small{}0.246}\tabularnewline
 & {\small{}250} & {\small{}0.900} & {\small{}0.252} & {\small{}0.108} & {\small{}0.040} & {\small{}0.174} & {\small{}0.048} & {\small{}0.904} & {\small{}0.233}\tabularnewline
 & {\small{}500} & {\small{}0.905} & {\small{}0.188} & {\small{}0.138} & {\small{}0.020} & {\small{}0.155} & {\small{}0.065} & {\small{}0.903} & {\small{}0.188}\tabularnewline
\hline 
{\small{}DGP2} & {\small{}5} & {\small{}0.926} & {\small{}2.474} & {\small{}0.661} & {\small{}0.232} & {\small{}0.845} & {\small{}0.283} & {\small{}0.867} & {\small{}0.331}\tabularnewline
 & {\small{}10} & {\small{}0.932} & {\small{}1.732} & {\small{}0.627} & {\small{}0.209} & {\small{}0.781} & {\small{}0.292} & {\small{}0.873} & {\small{}0.413}\tabularnewline
 & {\small{}20} & {\small{}0.922} & {\small{}2.261} & {\small{}0.591} & {\small{}0.261} & {\small{}0.677} & {\small{}0.252} & {\small{}0.876} & {\small{}0.395}\tabularnewline
 & {\small{}30} & {\small{}0.938} & {\small{}1.932} & {\small{}0.517} & {\small{}0.136} & {\small{}0.505} & {\small{}0.143} & {\small{}0.864} & {\small{}0.305}\tabularnewline
 & {\small{}40} & {\small{}0.934} & {\small{}2.351} & {\small{}0.472} & {\small{}0.170} & {\small{}0.305} & {\small{}0.122} & {\small{}0.879} & {\small{}0.344}\tabularnewline
 & {\small{}50} & {\small{}0.930} & {\small{}1.853} & {\small{}0.407} & {\small{}0.127} & {\small{}0.267} & {\small{}0.231} & {\small{}0.884} & {\small{}0.300}\tabularnewline
 & {\small{}100} & {\small{}0.918} & {\small{}2.063} & {\small{}0.250} & {\small{}0.069} & {\small{}0.566} & {\small{}2.065} & {\small{}0.876} & {\small{}0.570}\tabularnewline
 & {\small{}250} & {\small{}0.932} & {\small{}2.538} & {\small{}0.778} & {\small{}2.501} & {\small{}0.808} & {\small{}3.648} & {\small{}0.867} & {\small{}0.559}\tabularnewline
 & {\small{}500} & {\small{}0.915} & {\small{}2.286} & {\small{}0.881} & {\small{}6.466} & {\small{}0.875} & {\small{}5.146} & {\small{}0.876} & {\small{}0.494}\tabularnewline
\hline 
{\small{}DGP3} & {\small{}5} & {\small{}0.926} & {\small{}3.173} & {\small{}0.615} & {\small{}0.219} & {\small{}0.845} & {\small{}0.283} & {\small{}0.870} & {\small{}0.456}\tabularnewline
 & {\small{}10} & {\small{}0.932} & {\small{}2.243} & {\small{}0.601} & {\small{}0.202} & {\small{}0.781} & {\small{}0.292} & {\small{}0.905} & {\small{}0.724}\tabularnewline
 & {\small{}20} & {\small{}0.922} & {\small{}2.593} & {\small{}0.564} & {\small{}0.236} & {\small{}0.677} & {\small{}0.252} & {\small{}0.908} & {\small{}0.609}\tabularnewline
 & {\small{}30} & {\small{}0.931} & {\small{}2.498} & {\small{}0.495} & {\small{}0.151} & {\small{}0.505} & {\small{}0.143} & {\small{}0.881} & {\small{}0.680}\tabularnewline
 & {\small{}40} & {\small{}0.928} & {\small{}3.277} & {\small{}0.456} & {\small{}0.174} & {\small{}0.295} & {\small{}0.122} & {\small{}0.914} & {\small{}1.118}\tabularnewline
 & {\small{}50} & {\small{}0.927} & {\small{}2.462} & {\small{}0.397} & {\small{}0.132} & {\small{}0.189} & {\small{}0.082} & {\small{}0.888} & {\small{}0.444}\tabularnewline
 & {\small{}100} & {\small{}0.912} & {\small{}2.750} & {\small{}0.182} & {\small{}0.073} & {\small{}0.164} & {\small{}0.116} & {\small{}0.898} & {\small{}0.846}\tabularnewline
 & {\small{}250} & {\small{}0.931} & {\small{}3.578} & {\small{}0.119} & {\small{}0.139} & {\small{}0.175} & {\small{}0.362} & {\small{}0.895} & {\small{}0.745}\tabularnewline
 & {\small{}500} & {\small{}0.922} & {\small{}3.235} & {\small{}0.130} & {\small{}0.158} & {\small{}0.157} & {\small{}0.255} & {\small{}0.892} & {\small{}0.635}\tabularnewline
\hline 
{\small{}DGP4} & {\small{}5} & {\small{}0.910} & {\small{}1.083} & {\small{}0.807} & {\small{}0.269} & {\small{}0.845} & {\small{}0.283} & {\small{}0.899} & {\small{}0.525}\tabularnewline
 & {\small{}10} & {\small{}0.924} & {\small{}0.619} & {\small{}0.762} & {\small{}0.253} & {\small{}0.781} & {\small{}0.292} & {\small{}0.912} & {\small{}0.624}\tabularnewline
 & {\small{}20} & {\small{}0.923} & {\small{}0.948} & {\small{}0.689} & {\small{}0.277} & {\small{}0.677} & {\small{}0.252} & {\small{}0.904} & {\small{}0.854}\tabularnewline
 & {\small{}30} & {\small{}0.912} & {\small{}0.631} & {\small{}0.588} & {\small{}0.149} & {\small{}0.505} & {\small{}0.143} & {\small{}0.909} & {\small{}0.631}\tabularnewline
 & {\small{}40} & {\small{}0.932} & {\small{}0.993} & {\small{}0.489} & {\small{}0.162} & {\small{}0.296} & {\small{}0.122} & {\small{}0.925} & {\small{}0.997}\tabularnewline
 & {\small{}50} & {\small{}0.920} & {\small{}0.666} & {\small{}0.375} & {\small{}0.113} & {\small{}0.232} & {\small{}0.121} & {\small{}0.914} & {\small{}0.673}\tabularnewline
 & {\small{}100} & {\small{}0.914} & {\small{}0.714} & {\small{}0.306} & {\small{}0.289} & {\small{}0.471} & {\small{}0.835} & {\small{}0.914} & {\small{}0.714}\tabularnewline
 & {\small{}250} & {\small{}0.916} & {\small{}0.918} & {\small{}0.677} & {\small{}1.547} & {\small{}0.705} & {\small{}1.436} & {\small{}0.920} & {\small{}0.882}\tabularnewline
 & {\small{}500} & {\small{}0.910} & {\small{}0.821} & {\small{}0.853} & {\small{}1.862} & {\small{}0.791} & {\small{}3.522} & {\small{}0.910} & {\small{}0.810}\tabularnewline
\hline 
\end{tabular}{\small\par}
\end{table}

\newpage{}

\begin{table}[htp]
{\small{}\caption{{\small{}\label{tab:band}}Simulation: MSE-optimal bandwidths}
}{\small\par}
\centering{}{\small{}}%
\begin{tabular}{lccccccc}
\hline 
\multicolumn{2}{c}{{\small{}MSE-optimal bandwidths:}} & \multicolumn{2}{c}{{\small{}w/o Covariates}} & \multicolumn{2}{c}{{\small{}w/ Covariates}} & \multicolumn{2}{c}{{\small{}Adaptive}}\tabularnewline
 & {\small{}$p$} & {\small{}Mean} & {\small{}SD} & {\small{}Mean} & {\small{}SD} & {\small{}Mean} & {\small{}SD}\tabularnewline
\hline 
{\small{}DGP1} & {\small{}5} & {\small{}0.198} & {\small{}0.046} & {\small{}0.190} & {\small{}0.043} & {\small{}0.197} & {\small{}0.046}\tabularnewline
 & {\small{}10} & {\small{}0.195} & {\small{}0.046} & {\small{}0.179} & {\small{}0.041} & {\small{}0.195} & {\small{}0.046}\tabularnewline
 & {\small{}20} & {\small{}0.197} & {\small{}0.043} & {\small{}0.163} & {\small{}0.035} & {\small{}0.197} & {\small{}0.044}\tabularnewline
 & {\small{}30} & {\small{}0.196} & {\small{}0.044} & {\small{}0.147} & {\small{}0.031} & {\small{}0.196} & {\small{}0.045}\tabularnewline
 & {\small{}40} & {\small{}0.197} & {\small{}0.044} & {\small{}0.128} & {\small{}0.026} & {\small{}0.196} & {\small{}0.043}\tabularnewline
 & {\small{}50} & {\small{}0.194} & {\small{}0.045} & {\small{}0.108} & {\small{}0.024} & {\small{}0.193} & {\small{}0.045}\tabularnewline
 & {\small{}100} & {\small{}0.196} & {\small{}0.047} & {\small{}0.075} & {\small{}0.024} & {\small{}0.196} & {\small{}0.047}\tabularnewline
 & {\small{}250} & {\small{}0.194} & {\small{}0.045} & {\small{}0.069} & {\small{}0.019} & {\small{}0.194} & {\small{}0.045}\tabularnewline
 & {\small{}500} & {\small{}0.197} & {\small{}0.043} & {\small{}0.071} & {\small{}0.020} & {\small{}0.196} & {\small{}0.043}\tabularnewline
\hline 
{\small{}DGP2} & {\small{}5} & {\small{}0.177} & {\small{}0.025} & {\small{}0.108} & {\small{}0.011} & {\small{}0.111} & {\small{}0.013}\tabularnewline
 & {\small{}10} & {\small{}0.177} & {\small{}0.025} & {\small{}0.107} & {\small{}0.011} & {\small{}0.113} & {\small{}0.014}\tabularnewline
 & {\small{}20} & {\small{}0.176} & {\small{}0.026} & {\small{}0.105} & {\small{}0.011} & {\small{}0.115} & {\small{}0.015}\tabularnewline
 & {\small{}30} & {\small{}0.177} & {\small{}0.025} & {\small{}0.102} & {\small{}0.011} & {\small{}0.115} & {\small{}0.015}\tabularnewline
 & {\small{}40} & {\small{}0.176} & {\small{}0.025} & {\small{}0.099} & {\small{}0.012} & {\small{}0.117} & {\small{}0.015}\tabularnewline
 & {\small{}50} & {\small{}0.176} & {\small{}0.026} & {\small{}0.094} & {\small{}0.014} & {\small{}0.116} & {\small{}0.015}\tabularnewline
 & {\small{}100} & {\small{}0.176} & {\small{}0.025} & {\small{}0.130} & {\small{}0.026} & {\small{}0.118} & {\small{}0.016}\tabularnewline
 & {\small{}250} & {\small{}0.176} & {\small{}0.025} & {\small{}0.159} & {\small{}0.029} & {\small{}0.121} & {\small{}0.016}\tabularnewline
 & {\small{}500} & {\small{}0.177} & {\small{}0.025} & {\small{}0.173} & {\small{}0.031} & {\small{}0.121} & {\small{}0.016}\tabularnewline
\hline 
{\small{}DGP3} & {\small{}5} & {\small{}0.182} & {\small{}0.028} & {\small{}0.108} & {\small{}0.011} & {\small{}0.117} & {\small{}0.014}\tabularnewline
 & {\small{}10} & {\small{}0.183} & {\small{}0.029} & {\small{}0.107} & {\small{}0.011} & {\small{}0.138} & {\small{}0.015}\tabularnewline
 & {\small{}20} & {\small{}0.181} & {\small{}0.029} & {\small{}0.105} & {\small{}0.011} & {\small{}0.139} & {\small{}0.017}\tabularnewline
 & {\small{}30} & {\small{}0.183} & {\small{}0.028} & {\small{}0.102} & {\small{}0.011} & {\small{}0.138} & {\small{}0.015}\tabularnewline
 & {\small{}40} & {\small{}0.182} & {\small{}0.028} & {\small{}0.099} & {\small{}0.012} & {\small{}0.139} & {\small{}0.016}\tabularnewline
 & {\small{}50} & {\small{}0.183} & {\small{}0.029} & {\small{}0.092} & {\small{}0.013} & {\small{}0.140} & {\small{}0.015}\tabularnewline
 & {\small{}100} & {\small{}0.183} & {\small{}0.029} & {\small{}0.078} & {\small{}0.021} & {\small{}0.141} & {\small{}0.016}\tabularnewline
 & {\small{}250} & {\small{}0.181} & {\small{}0.028} & {\small{}0.076} & {\small{}0.022} & {\small{}0.142} & {\small{}0.018}\tabularnewline
 & {\small{}500} & {\small{}0.183} & {\small{}0.029} & {\small{}0.074} & {\small{}0.021} & {\small{}0.143} & {\small{}0.017}\tabularnewline
\hline 
{\small{}DGP4} & {\small{}5} & {\small{}0.141} & {\small{}0.015} & {\small{}0.108} & {\small{}0.011} & {\small{}0.126} & {\small{}0.016}\tabularnewline
 & {\small{}10} & {\small{}0.144} & {\small{}0.015} & {\small{}0.107} & {\small{}0.011} & {\small{}0.142} & {\small{}0.015}\tabularnewline
 & {\small{}20} & {\small{}0.144} & {\small{}0.016} & {\small{}0.105} & {\small{}0.011} & {\small{}0.142} & {\small{}0.016}\tabularnewline
 & {\small{}30} & {\small{}0.144} & {\small{}0.015} & {\small{}0.102} & {\small{}0.011} & {\small{}0.143} & {\small{}0.015}\tabularnewline
 & {\small{}40} & {\small{}0.145} & {\small{}0.015} & {\small{}0.099} & {\small{}0.012} & {\small{}0.144} & {\small{}0.016}\tabularnewline
 & {\small{}50} & {\small{}0.144} & {\small{}0.016} & {\small{}0.092} & {\small{}0.013} & {\small{}0.143} & {\small{}0.016}\tabularnewline
 & {\small{}100} & {\small{}0.145} & {\small{}0.015} & {\small{}0.104} & {\small{}0.021} & {\small{}0.143} & {\small{}0.016}\tabularnewline
 & {\small{}250} & {\small{}0.144} & {\small{}0.016} & {\small{}0.132} & {\small{}0.025} & {\small{}0.143} & {\small{}0.016}\tabularnewline
 & {\small{}500} & {\small{}0.144} & {\small{}0.015} & {\small{}0.150} & {\small{}0.025} & {\small{}0.143} & {\small{}0.016}\tabularnewline
\hline 
\end{tabular}{\small\par}
\end{table}

\newpage{}

\section{Empirical illustration: Head Start data\label{sec:emp}}

To illustrate our variable selection approach, we revisit the problem
of the Head Start program first studied by Ludwig and Miller (2007)
where they investigate the effect of the Head Start program on various
outcomes related to health and schooling. The federal government provided
grant-writing assistance to the 300 poorest counties based on the
poverty index to apply for the Head Start program. This leads to the
RDD with the poverty index as a running variable where the cut-off
value is set as $\bar{x}=59.1984$. Ludwig and Miller (2007) conducted
their RDD analysis using no covariate, and CCFT examined the impact
of the covariance-adjustment. CCFT employed nine pre-intervention
covariates from the U.S. Census, which include total population, percentage
of population, percentages of black and urban population, and levels
and percentages of population in three age groups (children aged 3
to 5, children aged 14 to 17, and adults older than 25). The main
finding by CCFT is that the covariate adjusted RDD inference yields
shorter confidence intervals while the RDD point estimates remain
stable.

An important aspect of the Head start example is that it is unclear
which covariates become useful to improve efficiency mainly due to
the lack of economic theories behind the problem. We conduct the empirical
exercises of CCFT by applying our variable selection approach with
two extensions. First, we introduce 36 interaction terms in addition
to the nine original covariates. Second, we also implement those estimation
and inference for subsamples to see the effect of changes in the ratio
of the number of covariates ($p$) to that of observations ($n$).
Hereafter, as in CCFT, we focus on child mortality among many outcome
variables.

Table 5 shows the results of our empirical illustration. Four columns
correspond to four estimation procedures which are the same as those
used in the simulation experiments. The first panel shows the full
sample results ($n=2799$ and $p/n=0.016$). The RDD causal effect
estimates are presented in the first row. The next three rows show
95\% confidence intervals, their percentage length changes relative
to the one in the first column, and their associated $p$-values where
these are obtained without restriction on the MSE optimal bandwidth
for the local linear regression ($h$) and the pilot bandwidth ($b$).
See CCT and CCFT for more details on the robust inference methods.
These results are also obtained under the restriction $h/b=1$, which
are reported in the following three rows. The next two rows in the
same panel present the bandwidths $(h,b)$, and effective sample sizes
$(n_{-},n_{+})$ used for the RDD estimation. The effective sample
sizes are the numbers of observations of the running variable in the
intervals $[\bar{x}-h,\bar{x}]$ and $[\bar{x},\bar{x}+h]$. We also
report the selected covariates for our covariate selection approach.
We use subsamples of the first 1000 and 500 observations for the second
and third panels, leading to $p/n=0.045$ and $0.090$, respectively. 

For the full sample case, the covariate adjusted estimates mildly
deviate from the standard one while our estimate based on the variable
selection is identical to the standard one. Although the confidence
intervals of the covariate adjusted approaches are shorter than the
standard one, this might induce under-coverages for the case of many
covariates as illustrated in the simulation experiment. As the sample
size gets smaller, the observations made here are amplified. In contrast,
we can see the stable performance of the variable selection approach
and its mild contribution to shorten the confidence intervals. Table
6 presents the corresponding results for the case of nine covariates
as considered by CCFT. The results are essentially similar to those
of Table 5, although the influence of the covariates is less dramatic.

\newpage{}

{\footnotesize{}}
\begin{table}[H]
{\footnotesize{}\caption{Empirical illustration: Head Start data ($45$ covariates)}
}{\footnotesize\par}
\centering{}{\footnotesize{}}%
\begin{tabular}{llcccc}
\hline 
 & {\footnotesize{}MSE-optimal bandwidths:} & \multicolumn{2}{c}{{\footnotesize{}w/o Covariates}} & {\footnotesize{}w/ Covariates} & {\footnotesize{}Adaptive}\tabularnewline
\hline 
 & {\footnotesize{}Estimation methods:} & {\footnotesize{}Standard} & {\footnotesize{}Cov-adjusted} & {\footnotesize{}Cov-adjusted} & {\footnotesize{}Variable selection}\tabularnewline
\hline 
{\footnotesize{}$n=2779$} & {\footnotesize{}Point estimate} & {\footnotesize{}$-2.41$} & {\footnotesize{}$-1.82$} & {\footnotesize{}$-3.64$} & {\footnotesize{}$-2.41$}\tabularnewline
{\footnotesize{}$p/n=0.016$} & {\footnotesize{}$h/b$ unrestricted} &  &  &  & \tabularnewline
 & {\footnotesize{}\hspace{3mm}Robust 95\% CI} & {\footnotesize{}$[-5.46,-0.1]$} & {\footnotesize{}$[-3.85,-0.07]$} & {\footnotesize{}$[-6.05,-1.24]$} & {\footnotesize{}$[-5.46,-0.1]$}\tabularnewline
 & {\footnotesize{}\hspace{3mm}CI length change (\%)} &  & {\footnotesize{}$-29.57$} & {\footnotesize{}$-10.31$} & {\footnotesize{}$0$}\tabularnewline
 & {\footnotesize{}\hspace{3mm}Robust p-value} & {\footnotesize{}$0.042$} & {\footnotesize{}$0.042$} & {\footnotesize{}$0.003$} & {\footnotesize{}$0.042$}\tabularnewline
 & {\footnotesize{}$h/b=1$} &  &  &  & \tabularnewline
 & {\footnotesize{}\hspace{3mm}Robust 95\% CI} & {\footnotesize{}$[-6.41,-1.09]$} & {\footnotesize{}$[-5.44,-1.10]$} & {\footnotesize{}$[-6.55,-1.14]$} & {\footnotesize{}$[-6.41,-1.09]$}\tabularnewline
 & {\footnotesize{}\hspace{3mm}CI length change (\%)} &  & {\footnotesize{}$-18.37$} & {\footnotesize{}$1.64$} & {\footnotesize{}$0$}\tabularnewline
 & {\footnotesize{}\hspace{3mm}Robust p-value} & {\footnotesize{}0.006} & {\footnotesize{}0.003} & {\footnotesize{}0.005} & {\footnotesize{}0.006}\tabularnewline
 & {\footnotesize{}$h$, $b$} & {\footnotesize{}6.81, 10.73} & {\footnotesize{}6.81, 10.73} & {\footnotesize{}3.02, 5.40} & {\footnotesize{}6.81, 10.73}\tabularnewline
 & {\footnotesize{}$n_{-}$, $n_{+}$} & {\footnotesize{}234, 180} & {\footnotesize{}234, 180} & {\footnotesize{}96, 84} & {\footnotesize{}234, 180}\tabularnewline
 & {\footnotesize{}Selected covariates} &  &  &  & {\footnotesize{}None}\tabularnewline
\hline 
{\footnotesize{}$n=1000$} & {\footnotesize{}Point estimate} & {\footnotesize{}$-1.68$} & {\footnotesize{}$-2.40$} & {\footnotesize{}$-3.44$} & {\footnotesize{}$-1.48$}\tabularnewline
{\footnotesize{}$p/n=0.045$} & {\footnotesize{}$h/b$ unrestricted} &  &  &  & \tabularnewline
 & {\footnotesize{}\hspace{3mm}Robust 95\% CI} & {\footnotesize{}$[-5.45,1.75]$} & {\footnotesize{}$[-5.44,-0.29]$} & {\footnotesize{}$[-6.14,-1.34]$} & {\footnotesize{}$[-5.08,1.79]$}\tabularnewline
 & {\footnotesize{}\hspace{3mm}CI length change (\%)} &  & {\footnotesize{}$-28.38$} & {\footnotesize{}$-33.23$} & {\footnotesize{}$-4.49$}\tabularnewline
 & {\footnotesize{}\hspace{3mm}Robust p-value} & {\footnotesize{}0.314} & {\footnotesize{}0.029} & {\footnotesize{}0.002} & {\footnotesize{}0.347}\tabularnewline
 & {\footnotesize{}$h/b=1$} &  &  &  & \tabularnewline
 & {\footnotesize{}\hspace{3mm}Robust 95\% CI} & {\footnotesize{}$[-8.26,0.22]$} & {\footnotesize{}$[-6.84,-0.47]$} & {\footnotesize{}$[-5.46,.0.15]$} & {\footnotesize{}$[-7.94,0.27]$}\tabularnewline
 & {\footnotesize{}\hspace{3mm}CI length change (\%)} &  & {\footnotesize{}$-24.95$} & {\footnotesize{}$-33.87$} & {\footnotesize{}$-3.28$}\tabularnewline
 & {\footnotesize{}\hspace{3mm}Robust p-value} & {\footnotesize{}0.063} & {\footnotesize{}0.025} & {\footnotesize{}0.064} & {\footnotesize{}0.070}\tabularnewline
 & {\footnotesize{}$h$, $b$} & {\footnotesize{}6.52, 10.23} & {\footnotesize{}6.52, 10.23} & {\footnotesize{}3.47, 6.00} & {\footnotesize{}5.26, 8.07}\tabularnewline
 & {\footnotesize{}$n_{-}$, $n_{+}$} & {\footnotesize{}74, 77} & {\footnotesize{}79, 79} & {\footnotesize{}40, 46} & {\footnotesize{}79, 79}\tabularnewline
 & {\footnotesize{}Selected covariates} &  &  &  & {\footnotesize{}\% of adult population}\tabularnewline
\hline 
{\footnotesize{}$n=500$} & {\footnotesize{}Point estimate} & {\footnotesize{}$-2.35$} & {\footnotesize{}$-4.23$} & {\footnotesize{}$-5.68$} & {\footnotesize{}$-2.22$}\tabularnewline
{\footnotesize{}$p/n=0.090$} & {\footnotesize{}$h/b$ unrestricted} &  &  &  & \tabularnewline
 & {\footnotesize{}\hspace{3mm}Robust 95\% CI} & {\footnotesize{}$[-7.25,2.48]$} & {\footnotesize{}$[-7.44,-2.06]$} & {\footnotesize{}$[-9.64,-4.63]$} & {\footnotesize{}$[-6.93,2.25]$}\tabularnewline
 & {\footnotesize{}\hspace{3mm}CI length change (\%)} &  & {\footnotesize{}$-44.77$} & {\footnotesize{}$-48.58$} & {\footnotesize{}$-5.74$}\tabularnewline
 & {\footnotesize{}\hspace{3mm}Robust p-value} & {\footnotesize{}0.337} & {\footnotesize{}0.001} & {\footnotesize{}0.000} & {\footnotesize{}0.317}\tabularnewline
 & {\footnotesize{}$h/b=1$} &  &  &  & \tabularnewline
 & {\footnotesize{}\hspace{3mm}Robust 95\% CI} & {\footnotesize{}$[-10.22,1.42]$} & {\footnotesize{}$[-9.03,-3.43]$} & {\footnotesize{}$[-9.62,-4.29]$} & {\footnotesize{}$[-10,1.07]$}\tabularnewline
 & {\footnotesize{}\hspace{3mm}CI length change (\%)} &  & {\footnotesize{}$-51.93$} & {\footnotesize{}$-54.16$} & {\footnotesize{}$-4.94$}\tabularnewline
 & {\footnotesize{}\hspace{3mm}Robust p-value} & {\footnotesize{}0.139} & {\footnotesize{}0.000} & {\footnotesize{}0.000} & {\footnotesize{}0.110}\tabularnewline
 & {\footnotesize{}$h$, $b$} & {\footnotesize{}6.37, 9.16} & {\footnotesize{}6.37, 9.16} & {\footnotesize{}4.96, 7.49} & {\footnotesize{}4.31, 6.82}\tabularnewline
 & {\footnotesize{}$n_{-}$, $n_{+}$} & {\footnotesize{}60, 56} & {\footnotesize{}61, 56} & {\footnotesize{}49, 47} & {\footnotesize{}61, 56}\tabularnewline
 & {\footnotesize{}Selected covariates} &  &  &  & {\footnotesize{}\% of adult population}\tabularnewline
\hline 
\end{tabular}{\footnotesize\par}
\end{table}
{\footnotesize\par}

{\footnotesize{}}
\begin{table}[H]
{\footnotesize{}\caption{Empirical illustration: Head Start data ($9$ covariates)}
}{\footnotesize\par}
\centering{}{\footnotesize{}}%
\begin{tabular}{llcccc}
\hline 
 & {\footnotesize{}MSE-optimal bandwidths:} & \multicolumn{2}{c}{{\footnotesize{}w/o Covariates}} & {\footnotesize{}w/ Covariates} & {\footnotesize{}Adaptive}\tabularnewline
\hline 
 & {\footnotesize{}Estimation methods:} & {\footnotesize{}Standard} & {\footnotesize{}Cov-adjusted} & {\footnotesize{}Cov-adjusted} & {\footnotesize{}Variable selection}\tabularnewline
\hline 
{\footnotesize{}$n=2779$} & {\footnotesize{}Point estimate} & {\footnotesize{}$-2.41$} & {\footnotesize{}$-2.51$} & {\footnotesize{}$2.47$} & {\footnotesize{}-2.41}\tabularnewline
{\footnotesize{}$p/n=0.003$} & {\footnotesize{}$h/b$ unrestricted} &  &  &  & \tabularnewline
 & {\footnotesize{}\hspace{3mm}Robust 95\% CI} & {\footnotesize{}$[-5.46,-0.1]$} & {\footnotesize{}$[-5.37,-0.45]$} & {\footnotesize{}$[-5.21,-0.37]$} & {\footnotesize{}$[-5.46,-0.1]$}\tabularnewline
 & {\footnotesize{}\hspace{3mm}CI length change (\%)} &  & {\footnotesize{}$-8.24$} & {\footnotesize{}$-9.74$} & {\footnotesize{}$0$}\tabularnewline
 & {\footnotesize{}\hspace{3mm}Robust p-value} & {\footnotesize{}$0.042$} & {\footnotesize{}$0.021$} & {\footnotesize{}$0.024$} & {\footnotesize{}$0.042$}\tabularnewline
 & {\footnotesize{}$h/b=1$} &  &  &  & \tabularnewline
 & {\footnotesize{}\hspace{3mm}Robust 95\% CI} & {\footnotesize{}$[-6.41,-1.09]$} & {\footnotesize{}$[-6.63,-1.46]$} & {\footnotesize{}$[-6.54,-1.39]$} & {\footnotesize{}$[-6.41,-1.09]$}\tabularnewline
 & {\footnotesize{}\hspace{3mm}CI length change (\%)} &  & {\footnotesize{}$-2.87$} & {\footnotesize{}$-3.23$} & {\footnotesize{}$0$}\tabularnewline
 & {\footnotesize{}\hspace{3mm}Robust p-value} & {\footnotesize{}0.006} & {\footnotesize{}0.002} & {\footnotesize{}0.003} & {\footnotesize{}0.006}\tabularnewline
 & {\footnotesize{}$h$, $b$} & {\footnotesize{}6.81, 10.73} & {\footnotesize{}6.81, 10.73} & {\footnotesize{}6.98, 11.64} & {\footnotesize{}6.81, 10.73}\tabularnewline
 & {\footnotesize{}$n_{-}$, $n_{+}$} & {\footnotesize{}234, 180} & {\footnotesize{}234, 180} & {\footnotesize{}240, 184} & {\footnotesize{}234, 180}\tabularnewline
 & {\footnotesize{}Selected covariates} &  &  &  & {\footnotesize{}None}\tabularnewline
\hline 
{\footnotesize{}$n=1000$} & {\footnotesize{}Point estimate} & {\footnotesize{}$-1.68$} & {\footnotesize{}$-2.05$} & {\footnotesize{}$-1.53$} & {\footnotesize{}$-1.48$}\tabularnewline
{\footnotesize{}$p/n=0.009$} & {\footnotesize{}$h/b$ unrestricted} &  &  &  & \tabularnewline
 & {\footnotesize{}\hspace{3mm}Robust 95\% CI} & {\footnotesize{}$[-5.45,1.75]$} & {\footnotesize{}$[-6.35,-1.1]$} & {\footnotesize{}$[-7.7,-2.28]$} & {\footnotesize{}$[-5.08,1.79]$}\tabularnewline
 & {\footnotesize{}\hspace{3mm}CI length change (\%)} &  & {\footnotesize{}$-7.12$} & {\footnotesize{}$-20.0$} & {\footnotesize{}$-4.49$}\tabularnewline
 & {\footnotesize{}\hspace{3mm}Robust p-value} & {\footnotesize{}0.314} & {\footnotesize{}0.141} & {\footnotesize{}0.237} & {\footnotesize{}0.347}\tabularnewline
 & {\footnotesize{}$h/b=1$} &  &  &  & \tabularnewline
 & {\footnotesize{}\hspace{3mm}Robust 95\% CI} & {\footnotesize{}$[-8.26,0.22]$} & {\footnotesize{}$[-8.26,-0.63]$} & {\footnotesize{}$[-5.93,-0.88]$} & {\footnotesize{}$[-7.94,0.27]$}\tabularnewline
 & {\footnotesize{}\hspace{3mm}CI length change (\%)} &  & {\footnotesize{}$-10.20$} & {\footnotesize{}$-19.75$} & {\footnotesize{}$-3.28$}\tabularnewline
 & {\footnotesize{}\hspace{3mm}Robust p-value} & {\footnotesize{}0.063} & {\footnotesize{}0.022} & {\footnotesize{}0.145} & {\footnotesize{}0.070}\tabularnewline
 & {\footnotesize{}$h$, $b$} & {\footnotesize{}6.52, 10.23} & {\footnotesize{}6.52, 10.23} & {\footnotesize{}9.14, 13.93} & {\footnotesize{}9.26, 8.07}\tabularnewline
 & {\footnotesize{}$n_{-}$, $n_{+}$} & {\footnotesize{}74, 77} & {\footnotesize{}79, 79} & {\footnotesize{}121, 98} & {\footnotesize{}79, 79}\tabularnewline
 & {\footnotesize{}Selected covariates} &  &  &  & {\footnotesize{}\% of adult population}\tabularnewline
\hline 
{\footnotesize{}$n=500$} & {\footnotesize{}Point estimate} & {\footnotesize{}$-2.35$} & {\footnotesize{}$-2.54$} & {\footnotesize{}$-2.39$} & {\footnotesize{}$-2.22$}\tabularnewline
{\footnotesize{}$p/n=0.018$} & {\footnotesize{}$h/b$ unrestricted} &  &  &  & \tabularnewline
 & {\footnotesize{}\hspace{3mm}Robust 95\% CI} & {\footnotesize{}$[-7.25,2.48]$} & {\footnotesize{}$[-7.25,1.28]$} & {\footnotesize{}$[-6.75,1.6]$} & {\footnotesize{}$[-6.93,2.25]$}\tabularnewline
 & {\footnotesize{}\hspace{3mm}CI length change (\%)} &  & {\footnotesize{}$-12.36$} & {\footnotesize{}$-14.24$} & {\footnotesize{}$-5.74$}\tabularnewline
 & {\footnotesize{}\hspace{3mm}Robust p-value} & {\footnotesize{}0.337} & {\footnotesize{}0.170} & {\footnotesize{}0.227} & {\footnotesize{}0.317}\tabularnewline
 & {\footnotesize{}$h/b=1$} &  &  &  & \tabularnewline
 & {\footnotesize{}\hspace{3mm}Robust 95\% CI} & {\footnotesize{}$[-10.22,1.42]$} & {\footnotesize{}$[-9.97,-0.03]$} & {\footnotesize{}$[-9.74,0.03]$} & {\footnotesize{}$[-10,1.07]$}\tabularnewline
 & {\footnotesize{}\hspace{3mm}CI length change (\%)} &  & {\footnotesize{}$-14.58$} & {\footnotesize{}$-16.10$} & {\footnotesize{}$-4.94$}\tabularnewline
 & {\footnotesize{}\hspace{3mm}Robust p-value} & {\footnotesize{}0.139} & {\footnotesize{}0.049} & {\footnotesize{}0.051} & {\footnotesize{}0.110}\tabularnewline
 & {\footnotesize{}$h$, $b$} & {\footnotesize{}6.37, 9.16} & {\footnotesize{}6.37, 9.16} & {\footnotesize{}6.58, 9.88} & {\footnotesize{}4.31, 6.82}\tabularnewline
 & {\footnotesize{}$n_{-}$, $n_{+}$} & {\footnotesize{}60, 56} & {\footnotesize{}61, 56} & {\footnotesize{}63, 58} & {\footnotesize{}61, 56}\tabularnewline
 & {\footnotesize{}Selected covariates} &  &  &  & {\footnotesize{}\% of adult population}\tabularnewline
\hline 
\end{tabular}{\footnotesize\par}
\end{table}
{\footnotesize\par}

\newpage{}

\begin{figure}[H]
	\caption{Comparison of three approaches ($n=1000$, $45$ covariates) for point
		estimates (left panel) and CI lengths (right panel) }
	
	\begin{centering}
		\includegraphics[scale=0.4]{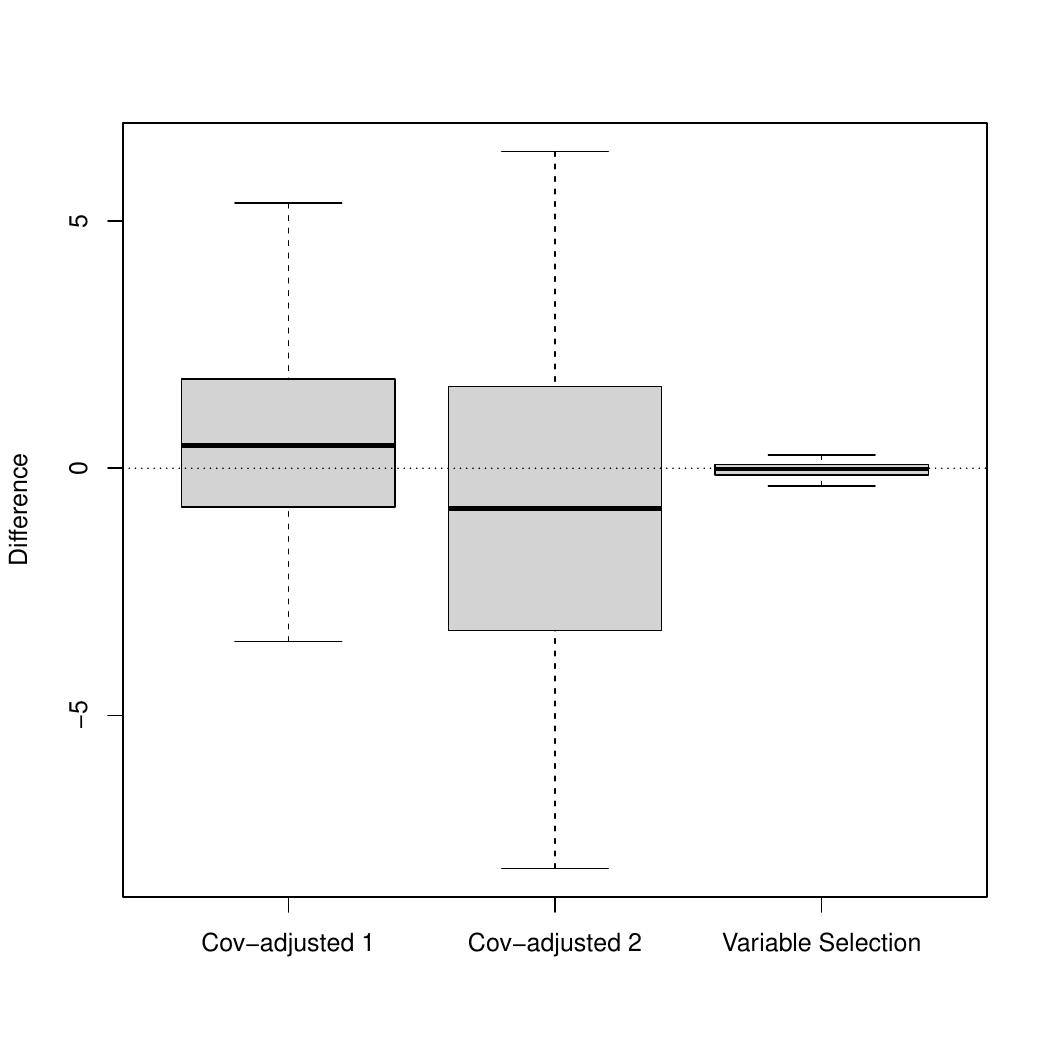}\includegraphics[scale=0.4]{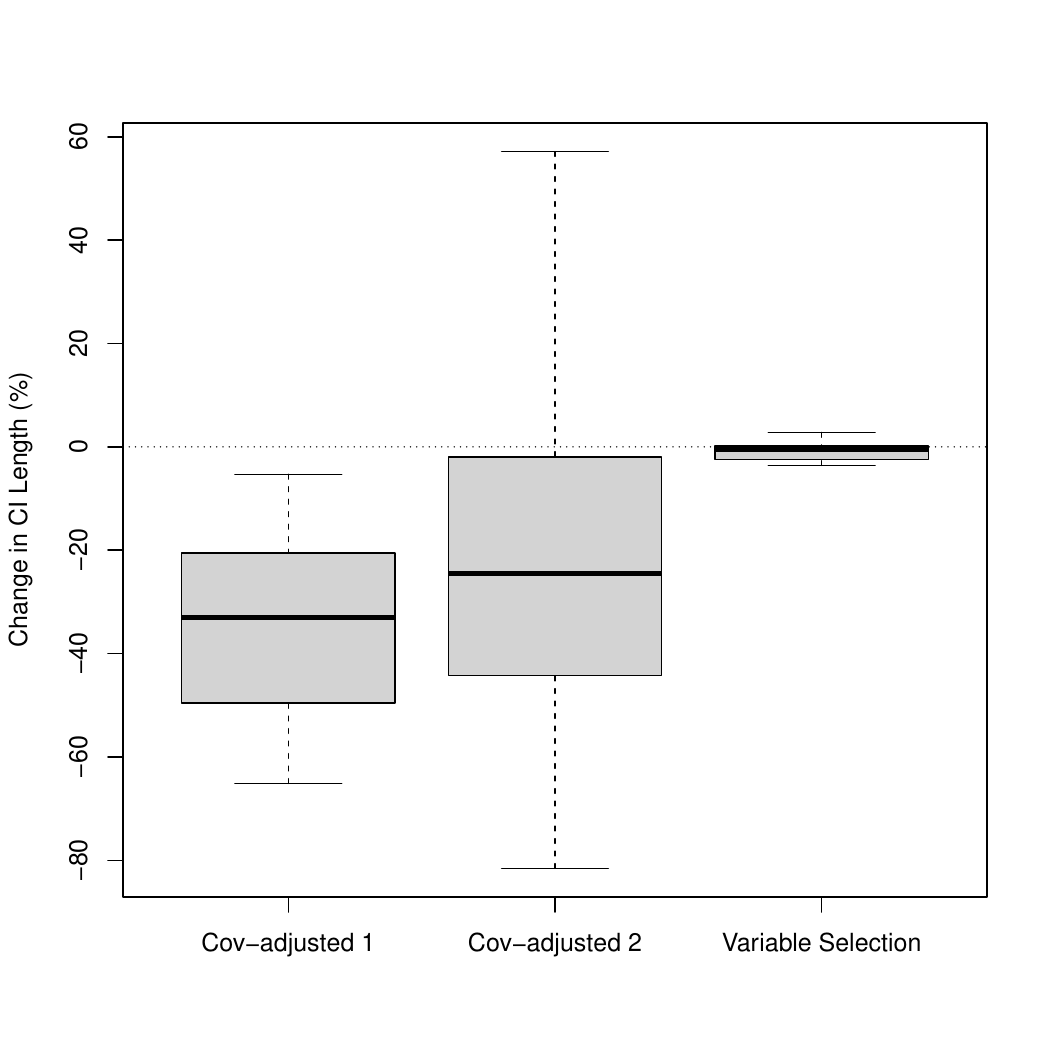}
		\par\end{centering}
	\centering{}{\footnotesize{}\label{fig:1}}{\footnotesize\par}
\end{figure}

\begin{figure}[H]
	\caption{Comparison of three approaches ($n=500$, 45 covariates) for point
		estimates (left panel) and CI lengths (right panel) }
	
	\begin{centering}
		\includegraphics[scale=0.4]{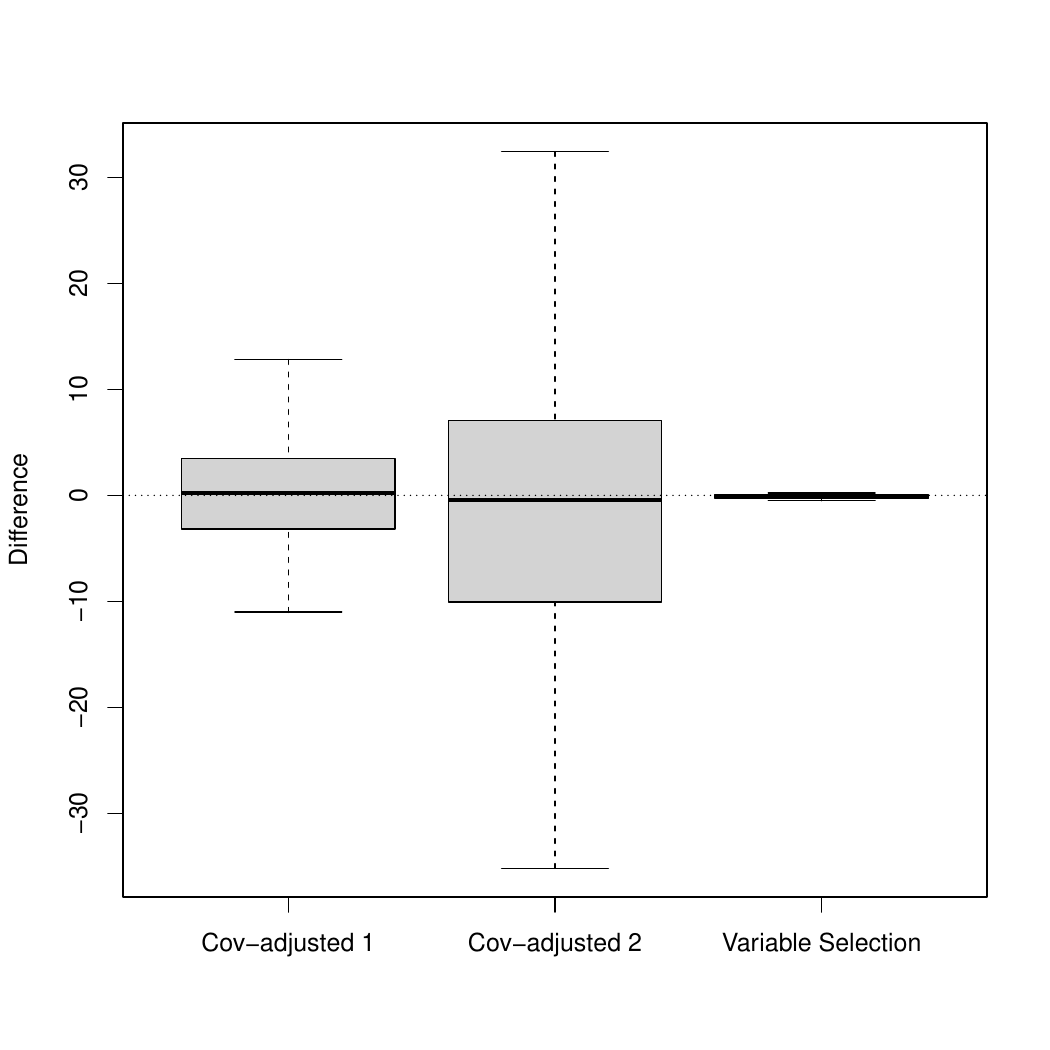}\includegraphics[scale=0.4]{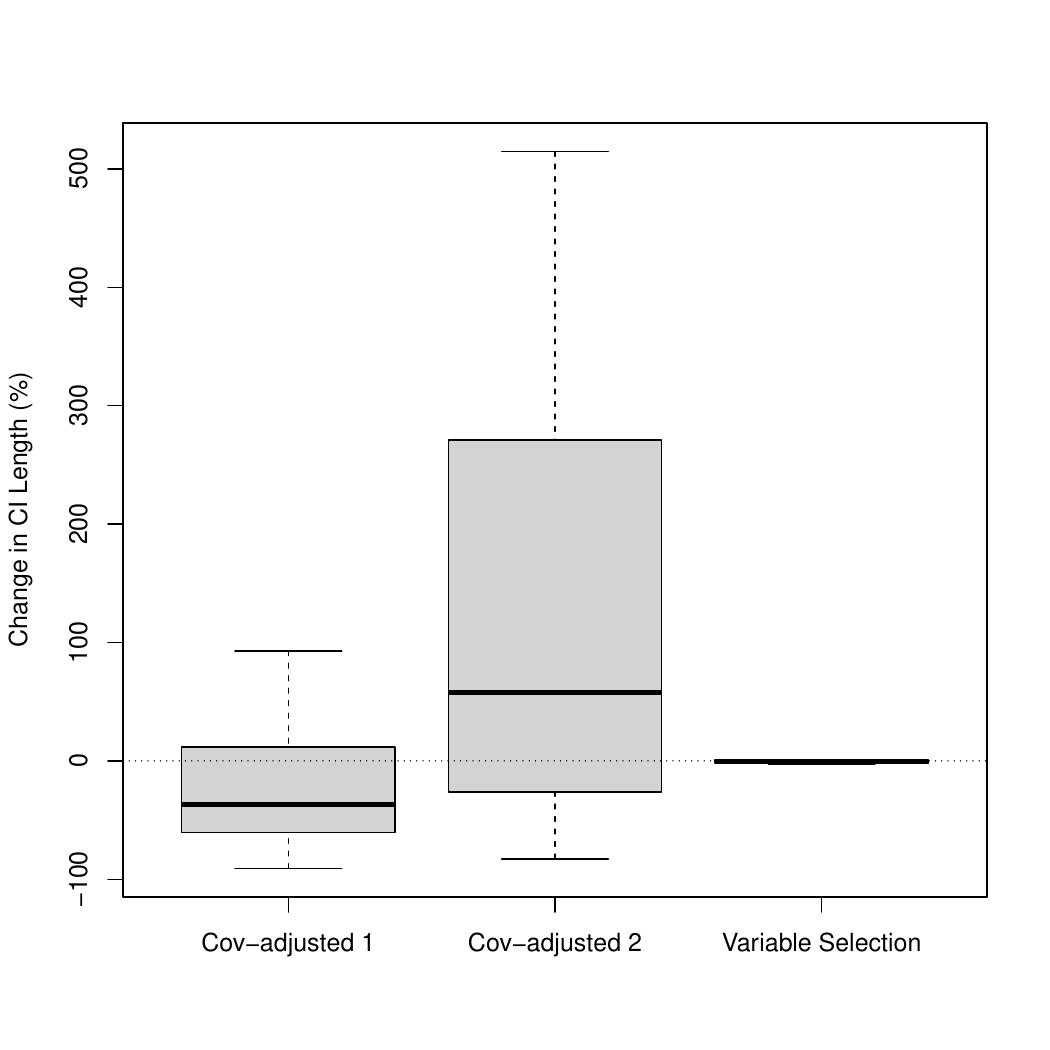}
		\par\end{centering}
	\centering{}{\footnotesize{}\label{fig:2}}{\footnotesize\par}
\end{figure}

\begin{figure}
	\caption{Comparison of three approaches ($n=1000$, 9 covariates) for point
		estimates (left panel) and CI lengths (right panel)}
	
	\begin{centering}
		\includegraphics[scale=0.4]{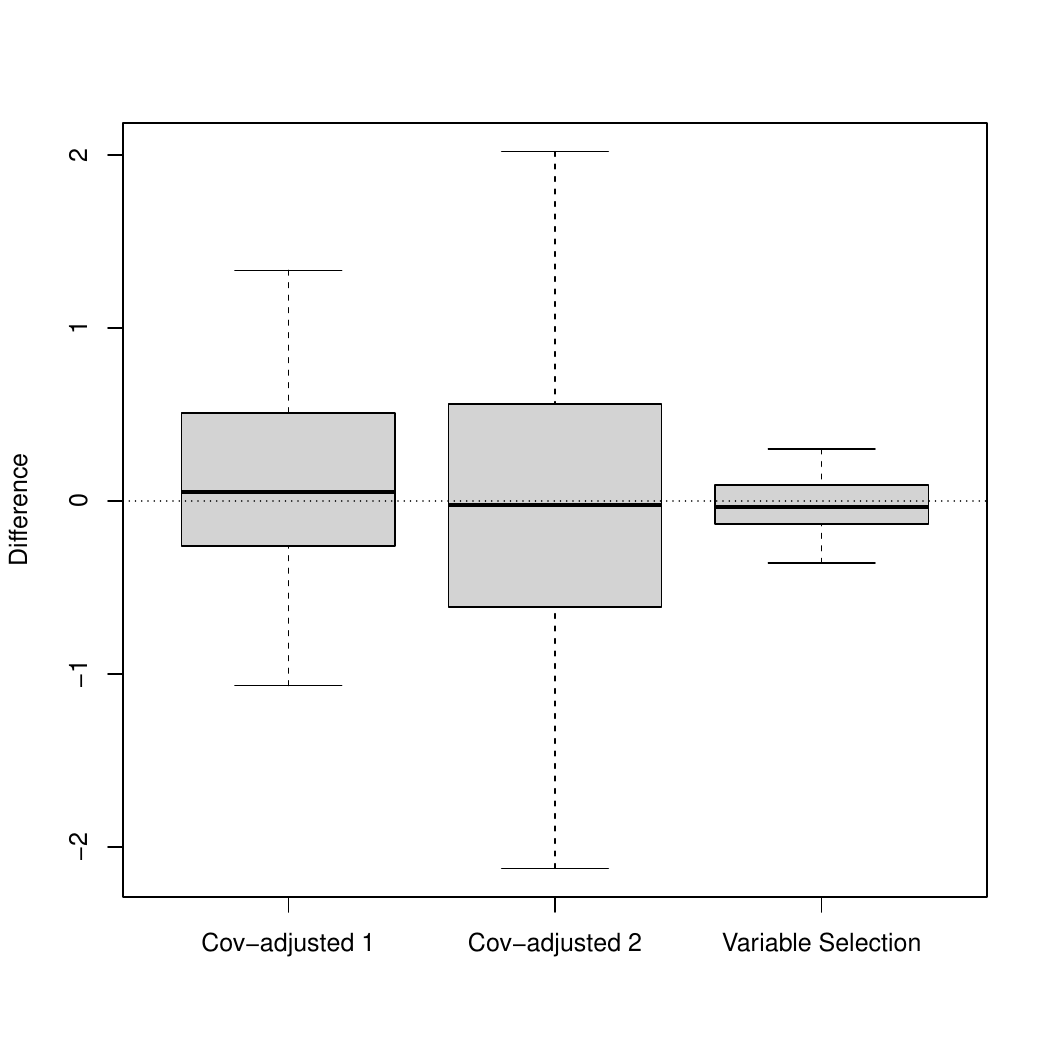}\includegraphics[scale=0.4]{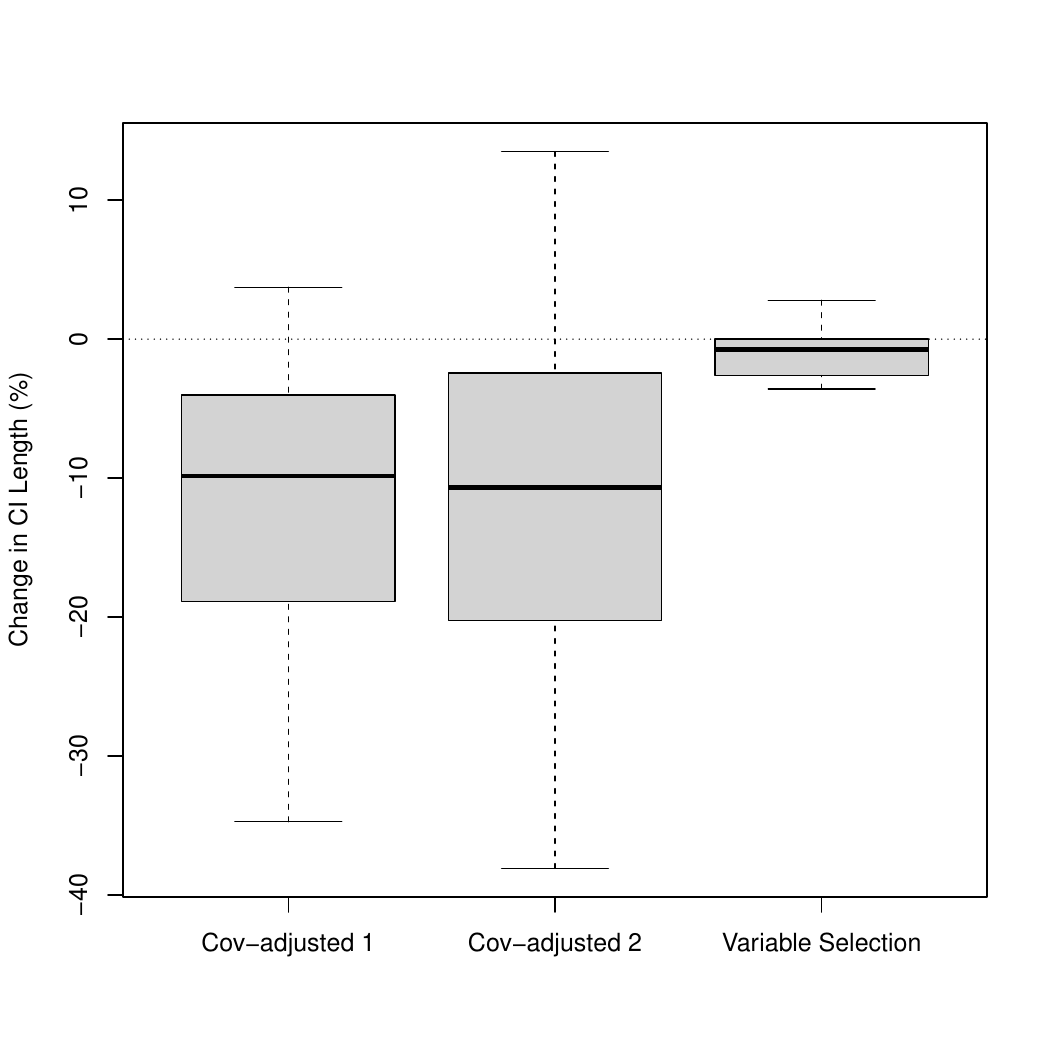}
		\par\end{centering}
	\centering{}{\footnotesize{}\label{fig:3}}{\footnotesize\par}
\end{figure}

\begin{figure}
	\caption{Comparison of three approaches ($n=500$, 9 covariates) for point
		estimates (left panel) and CI lengths (right panel)}
	
	\begin{centering}
		\includegraphics[scale=0.4]{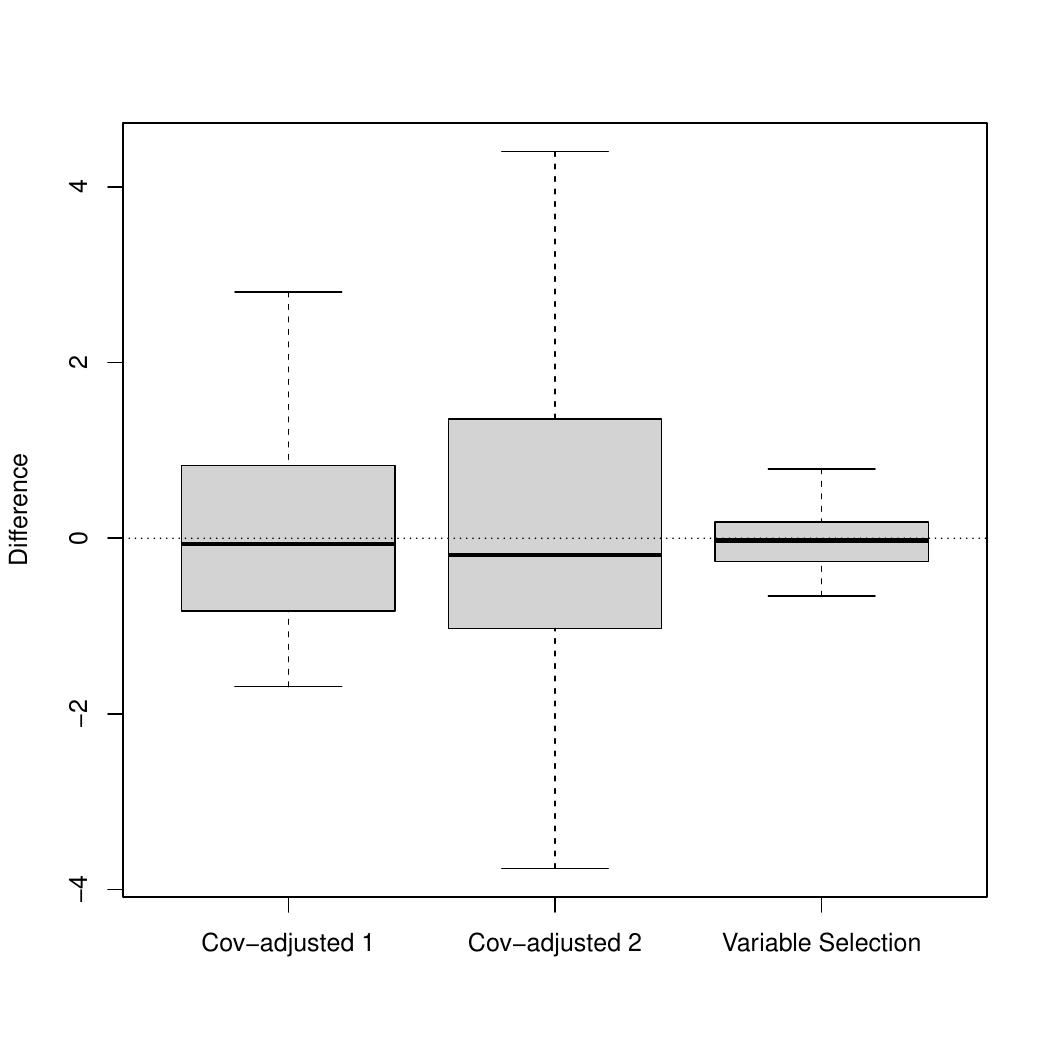}\includegraphics[scale=0.4]{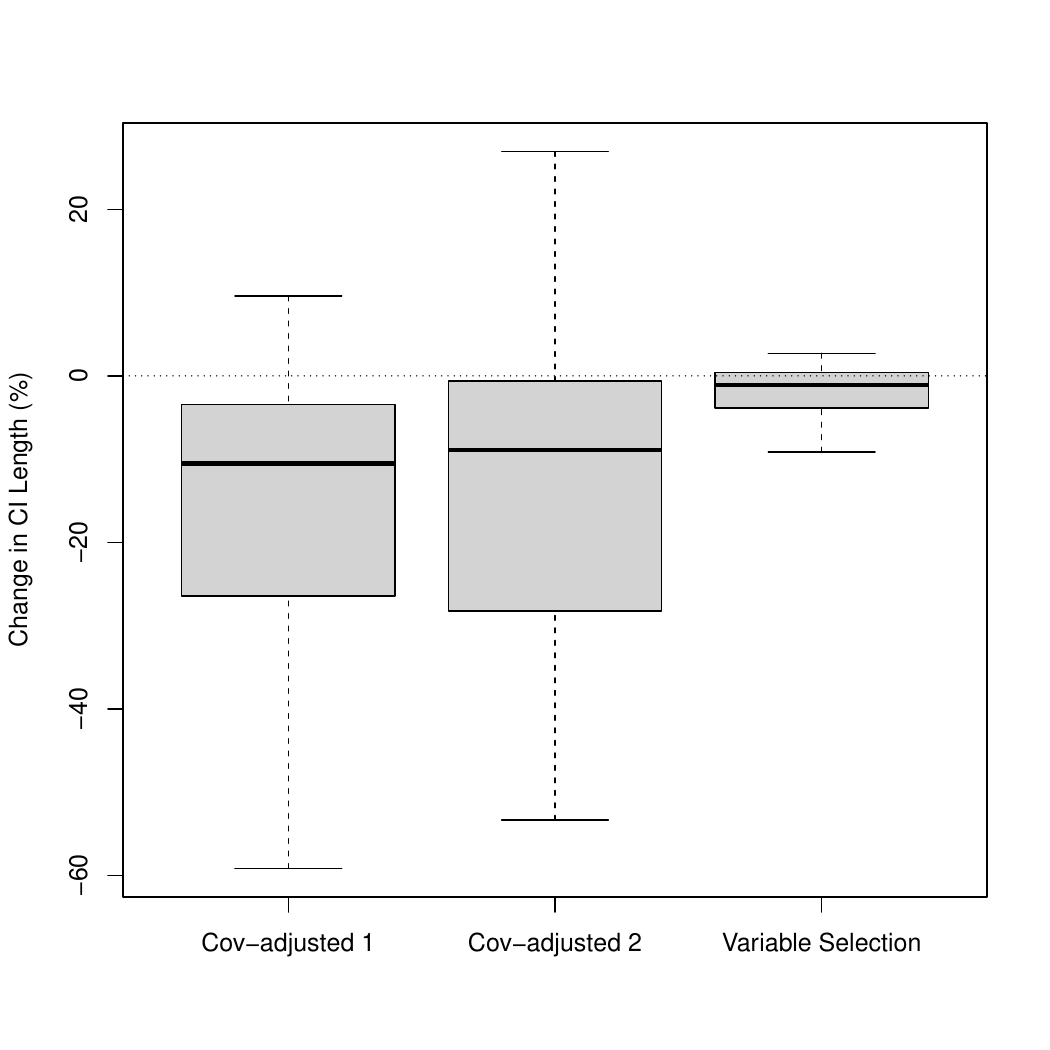}
		\par\end{centering}
	\centering{}{\footnotesize{}\label{fig:4}}{\footnotesize\par}
\end{figure}

To better understand the proposed method in this paper, we investigate
behaviors of three approaches, two covariate adjusted and variable
selection approaches, through the Head start example where the approaches
are the same as those considered in the simulation, and Tables 5 and
6. The approach denoted by ``Cov-adjusted 1'' corresponds to the covariate
adjusted approach using the CCT MSE-optimal bandwidths, and the one
denoted by ``Cov-adjusted 2'' to the one using the CCFT MSE-optimal
bandwidths. We construct 50 subsamples of $k$ observations out of
the Head start data where $k$ is set to 1000, and 500. We estimate
the RDD treatment effects based on the subsamples using nine or forty-five
covariates where nine covariates are those considered in CCFT and
forty-five are those considered in Table 5. Figures \ref{fig:1}-\ref{fig:4}
show boxplots of the results on estimation (left panel) and inference
(right panel).\footnote{The results shown in Figures \ref{fig:1}-\ref{fig:3} are based on
the bandwidths with $h_{n}/b_{n}$ unrestricted, and those with the
bandwidths with $h_{n}/b_{n}=1$ are qualitatively similar.} The top and bottom of the boxes are the third and first quartiles,
and the top- and bottom-bars show the maximum and minimum values less
than (the third quartiles + 1.5 times the interquartile range) and
greater than (the first quartile - 1.5 times the interquartile range),
respectively. The left panels in Figures \ref{fig:1}-\ref{fig:4}
show the differences in the RDD treatment effects between the three
approaches and the standard one where the standard one is based on
CCT. The right panel in each figure shows the corresponding differences
in the confidence interval lengths.\footnote{The variable selection approach produced the identical result to the
standard one about 20 times for all setups. The boxplots are drawn
based on the non-identical results.}

First, the left panels of Figures \ref{fig:1} and \ref{fig:2} show
the point estimates by the covariate adjusted approaches deviate from
the standard one by a large extent when the number of covariates is
large, and the differences get larger as the sample size becomes small.
We observe a similar tendency for the situations with nine covariates,
while differences are less dramatic. Second, we note that the point
estimates based on the standard and variable selection approaches
are very close, and they are stable. Third, we observe that the covariate
adjusted approaches tend to produce large decreases in the interval
length, which can reflect under-coverages as observed in Table \ref{tab:inf}.
Fourth, the variable selection approach leads to reductions of around
5\%, which are still nonnegligible. Fifth, the results for nine covariates
given in Figures \ref{fig:3} and \ref{fig:4} are similar to those
for forty-five covariates, although the magnitude of changes for nine
covariates is not as extreme as that for forty-five covariates. These
results show the usefulness of our variable selection approach not
only for the situation where the number of covariates is large but
also for that where the number of covariates is relatively small,
when the number of observations is not so large.

We often encounter situations where a number of potentially useful
covariates are available. Furthermore, It is common to employ transformations
of covariates such as interaction and quadratic terms. However, we
are typically uncertain about which covariates contribute to improving
efficiency possibly due to the lack of economic theories. The RDD
analysis is local in nature, and the number of covariates relative
to the effective sample size can be pretty large. We have seen that
covariate adjusted approaches can become misleading in several examples.
The steady performance of the variable selection approach under various
circumstances is noteworthy, and it can provide an essential second
opinion for the standard and covariate adjusted approaches.

\appendix

\section{Mathematical appendix}

\subsection{Proof of Theorem \ref{asm:point-est} \label{sub:pf1}}

We use the following modification of Bernstein's inequality.

\begin{lem}\label{lem:Bernstein} Under Assumptions \ref{asm:point-est}-\ref{asm:kb},
it holds
\[
\mathbb{E}\left[\max_{1\le j\le p}\left|\frac{1}{\sqrt{n}}\sum_{i=1}^{n}\left\{ K\left(\frac{X_{i}}{b_{n}}\right)G_{i,j}\epsilon_{i}-\mathbb{E}\left[K\left(\frac{X_{i}}{b_{n}}\right)G_{i,j}\epsilon_{i}\right]\right\} \right|^{m}\right]\leq2C^{m}b_{n}^{m/2}\log^{m/2}p,
\]
for $m\leq1+\log p$. \end{lem}

\textbf{Proof of Lemma \ref{lem:Bernstein}.} It follows from Bernstein's
inequality (e.g., Lemma 14.12 in Bühlmann and van de Geer, 2011) that
\[
\mathbb{E}\left[\max_{1\le j\le p}\left|\frac{1}{\sqrt{n}}\sum_{i=1}^{n}\left\{ K\left(\frac{X_{i}}{b_{n}}\right)G_{i,j}\epsilon_{i}-\mathbb{E}\left[K\left(\frac{X_{i}}{b_{n}}\right)G_{i,j}\epsilon_{i}\right]\right\} \right|^{m}\right]\leq2C^{m}b_{n}^{m/2}\log^{m/2}p,
\]
for $m\leq1+\log p$. $\square$

To make the proof more accessible and comparable to the standard Lasso,
we begin with $\tilde{\theta}$ as the solution of
\begin{equation}
\min_{\theta}\frac{1}{nb_{n}}\sum_{i=1}^{n}K\left(\frac{X_{i}}{b_{n}}\right)(Y_{i}-\alpha-T_{i}\tau-X_{i}\beta_{-}-T_{i}X_{i}\beta_{+}-Z_{i}^{\prime}\gamma)^{2}+\lambda_{n}|\theta|_{1}.\label{eq:lasso-2}
\end{equation}

\subsubsection{Deviation bounds for $\tilde{\theta}$}

Under Assumptions \ref{asm:point-est}-\ref{asm:kb}, Lemma \ref{lem:Bernstein}
implies 
\[
\mathbb{E}\left[\max_{1\le j\le p}\left|\frac{1}{\sqrt{n}}\sum_{i=1}^{n}K\left(\frac{X_{i}}{b_{n}}\right)G_{ij}\epsilon_{i}\right|\right]\leq2b_{n}^{1/2}\log^{1/2}p.
\]
Thus, we have 
\begin{equation}
\mathbb{P}\{\mathcal{A}_{n}\}:=\mathbb{P}\left\{ \frac{4}{nb_{n}}\left|\sum_{i=1}^{n}K\left(\frac{X_{i}}{b_{n}}\right)G_{i}\epsilon_{i}\right|_{\infty}\leq\lambda_{n}\right\} \rightarrow1,\label{pf:1-1}
\end{equation}
as $n\to\infty$, provided that $\sqrt{\log p/(nb_{n})}=o(\lambda_{n})$. 

Let $\mathbf{Y}=(Y_{1}K_{1}^{1/2},\ldots,Y_{n}K_{n}^{1/2})^{\prime}$,
$\mathbf{e}=(\epsilon_{1}K_{1}^{1/2},\ldots,\epsilon_{n}K_{n}^{1/2})^{\prime}$,
and $\mathbf{G}=(G_{1}K_{1}^{1/2},\ldots,G_{n}K_{n}^{1/2})^{\prime}$.
Since $\tilde{\theta}$ is a minimizer, we have 
\[
\frac{1}{nb_{n}}|\mathbf{Y}-\mathbf{G}\tilde{\theta}|_{2}^{2}+\lambda_{n}|\tilde{\theta}|_{1}\leq\frac{1}{nb_{n}}|\mathbf{Y}-\mathbf{G}\theta^{*}|_{2}^{2}+\lambda_{n}|\theta^{*}|_{1}.
\]
By plugging $\mathbf{Y}=\mathbf{G}\theta^{*}+\mathbf{e}$ into the
above, we obtain
\begin{eqnarray}
\frac{2}{nb_{n}}|\mathbf{G}(\tilde{\theta}-\theta^{*})|_{2}^{2} & \leq & \frac{4}{nb_{n}}\mathbf{e}^{\prime}\mathbf{G}(\tilde{\theta}-\theta^{*})+2\lambda_{n}|\theta^{*}|_{1}-2\lambda_{n}|\tilde{\theta}|_{1}\nonumber \\
 & \leq & \frac{4}{nb_{n}}|\mathbf{e}^{\prime}\mathbf{G}|_{\infty}|\tilde{\theta}-\theta^{*}|_{1}+2\lambda_{n}(|\theta^{*}|_{1}-|\tilde{\theta}|_{1})\nonumber \\
 & \leq & 3\lambda_{n}|\tilde{\theta}_{S^{*}}-\theta_{S^{*}}^{*}|_{1}-\lambda_{n}|\tilde{\theta}_{S_{c}^{*}}|_{1},\label{pf:basic}
\end{eqnarray}
conditionally on $\mathcal{A}_{n}$, where $S_{c}^{*}$ is the complement
of $S^{*}$, the second inequality follows from the Hölder inequality,
and the third inequality follows from the definition of $\mathcal{A}_{n}$
and the following facts
\begin{eqnarray}
|\tilde{\theta}-\theta^{*}|_{1} & = & |\tilde{\theta}_{S_{\ast}}-\theta_{S_{\ast}}^{\ast}|_{1}+|\tilde{\theta}_{S_{c}^{*}}|_{1},\label{pf:1-2}\\
|\theta^{\ast}|_{1}-|\tilde{\theta}|_{1} & = & |\theta_{S^{*}}^{\ast}|_{1}-|\tilde{\theta}_{S^{*}}|_{1}-|\tilde{\theta}_{S_{c}^{*}}|_{1}\leq|\tilde{\theta}_{S^{*}}-\theta_{S^{*}}^{\ast}|_{1}-|\tilde{\theta}_{S_{c}^{*}}|_{1},\nonumber 
\end{eqnarray}
due to the triangle inequality. Thus, (\ref{pf:basic}) implies $3|\tilde{\theta}_{S^{*}}-\theta_{S^{*}}^{\ast}|_{1}\geq|\tilde{\theta}_{S_{c}^{*}}|_{1}$
and
\begin{equation}
\frac{2}{nb_{n}}|\mathbf{G}(\tilde{\theta}-\theta^{*})|_{2}^{2}+\lambda_{n}|\tilde{\theta}-\theta^{*}|_{1}\leq4\lambda_{n}|\tilde{\theta}_{S^{*}}-\theta_{S^{*}}^{\ast}|_{1},\label{pf:1-3}
\end{equation}
by using (\ref{pf:1-2}).

Now, Assumption \ref{asm:point-est} implies that
\[
4\lambda_{n}|\tilde{\theta}_{S^{*}}-\theta_{S^{*}}^{\ast}|_{1}\leq4\lambda_{n}\sqrt{\frac{s^{*}}{nb_{n}\phi^{*2}}}|\mathbf{G}(\tilde{\theta}-\theta^{\ast})|_{2}\leq\frac{1}{nb_{n}}|\mathbf{G}(\tilde{\theta}-\theta^{\ast})|_{2}^{2}+4\lambda_{n}^{2}\frac{s^{*}}{\phi^{*2}},
\]
with probability approaching one, where we note that $2ab\leq a^{2}+b^{2}$
for the second inequality. Combining this with (\ref{pf:1-3}), we
have 
\[
\frac{1}{nb_{n}}|\mathbf{G}(\tilde{\theta}-\theta^{\ast})|_{2}^{2}+\lambda_{n}|\tilde{\theta}-\theta^{\ast}|_{1}\leq4\lambda_{n}^{2}\frac{s^{*}}{\phi^{*2}},
\]
with probability approaching one.

Also note that this result implies that the number of nonzero elements
in $\tilde{\theta}$ is $O(s^{*})$ by the argument given in Bickel
\emph{et al.} (2009, eq. (B.3)). 

\subsubsection{Proof of (i)}

Let $G_{i}=(G_{1i}^{\prime},G_{2i}^{\prime})^{\prime}$, where $G_{2i}=K_{i}^{1/2}Z_{i}$
and $G_{1i}$ is the other in the partition of $G_{i}$. Define $\mathbf{G}_{1}=(G_{11},\ldots,G_{1n})^{\prime}$
and $\mathbf{G}_{2}=(G_{21},\ldots,G_{2n})^{\prime}$ and partition
$\theta$ accordingly. The compatibility condition for $\mathbf{G}_{1}$
is implied by the usual full column rank condition in the classical
linear regression since $|\theta_{1}|_{1}^{2}\leq\dim(\theta_{1})|\theta_{1}|_{2}^{2}$.

Note that the result in (\ref{pf:1-1}) still holds. Since $\hat{\theta}$
is a minimizer, we have 
\[
\frac{1}{nb_{n}}|\mathbf{Y}-\mathbf{G}\hat{\theta}|_{2}^{2}+\lambda_{n}|\hat{\theta}_{2}|_{1}\leq\frac{1}{nb_{n}}|\mathbf{Y}-\mathbf{G}\theta^{*}|_{2}^{2}+\lambda_{n}|\theta_{2}^{*}|_{1}.
\]
By plugging $\mathbf{Y}=\mathbf{G}\theta^{*}+\mathbf{e}$ into the
above, we have
\begin{eqnarray}
\frac{2}{nb_{n}}|\mathbf{G}(\hat{\theta}-\theta^{\ast})|_{2}^{2} & \leq & \frac{4}{nb_{n}}\mathbf{e}^{\prime}\mathbf{G}(\hat{\theta}-\theta^{\ast})+2\lambda_{n}|\theta_{2}^{\ast}|_{1}-2\lambda_{n}|\hat{\theta}_{2}|_{1}\nonumber \\
 & \leq & \frac{4}{nb_{n}}|\mathbf{e}^{\prime}\mathbf{G}|_{\infty}|\hat{\theta}-\theta^{\ast}|_{1}+2\lambda_{n}(|\theta_{2}^{\ast}|_{1}-|\hat{\theta}_{2}|_{1})\nonumber \\
 & \leq & \lambda_{n}|\hat{\theta}_{1}-\theta_{1}^{*}|_{1}+3\lambda_{n}|\hat{\theta}_{2,S^{*}}-\theta_{2,S^{*}}^{\ast}|_{1}-\lambda_{n}|\hat{\theta}_{2,S_{c}^{*}}|_{1},\label{pf:basic2}
\end{eqnarray}
conditionally on $\mathcal{A}_{n}$, where the second inequality follows
from the Hölder inequality, and the third inequality follows from
the definition of $\mathcal{A}_{n}$ and the following facts
\begin{eqnarray}
|\hat{\theta}-\theta^{\ast}|_{1} & = & |\hat{\theta}_{S^{*}}-\theta_{S^{*}}^{\ast}|_{1}+|\hat{\theta}_{S_{c}^{*}}|_{1},\label{pf:2-1}\\
|\theta^{\ast}|_{1}-|\hat{\theta}|_{1} & = & |\theta_{S^{*}}^{\ast}|_{1}-|\hat{\theta}_{S^{*}}|_{1}-|\hat{\theta}_{S_{c}^{*}}|_{1}\leq|\hat{\theta}_{S^{*}}-\theta_{S^{*}}^{\ast}|_{1}-|\hat{\theta}_{S_{c}^{*}}|_{1},\nonumber 
\end{eqnarray}
due to the triangle inequality. Thus, (\ref{pf:basic2}) implies $3|\hat{\theta}_{S_{\ast}}-\theta_{S_{\ast}}^{\ast}|_{1}\geq|\hat{\theta}_{S_{\ast}^{c}}|_{1}$
and 
\begin{equation}
\frac{2}{nb_{n}}|\mathbf{G}(\hat{\theta}-\theta^{\ast})|_{2}^{2}+\lambda_{n}|\hat{\theta}_{2}-\theta_{2}^{\ast}|_{1}\leq\lambda_{n}|\hat{\theta}_{1}-\theta_{1}^{*}|_{1}+4\lambda_{n}|\hat{\theta}_{2,S^{*}}-\theta_{2,S^{*}}^{\ast}|_{1},\label{pf:2-2}
\end{equation}
by using (\ref{pf:2-1}).

Now Assumption \ref{asm:point-est} implies
\begin{eqnarray*}
4\lambda_{n}|\hat{\theta}_{2,S^{*}}-\theta_{2,S^{*}}^{\ast}|_{1} & \leq & 4\lambda_{n}\sqrt{\frac{s^{*}}{nb_{n}\phi^{*2}}}|\mathbf{G}_{2}(\hat{\theta}_{2}-\theta_{2}^{\ast})|_{2}\leq\frac{1}{nb_{n}}|\mathbf{G}_{2}(\hat{\theta}_{2}-\theta_{2}^{\ast})|_{2}^{2}+4\lambda_{n}^{2}\frac{s^{*}}{\phi^{*2}},\\
\lambda_{n}|\hat{\theta}_{1}-\theta_{1}^{*}|_{1} & \le & \lambda_{n}\sqrt{\frac{s^{*}}{nb_{n}\phi^{*2}}}|\mathbf{G}_{1}(\hat{\theta}_{1}-\theta_{1}^{\ast})|_{2}\leq\frac{1}{nb_{n}}|\mathbf{G}_{1}(\hat{\theta}_{1}-\theta_{1}^{\ast})|_{2}^{2}+\lambda_{n}^{2}\frac{s^{*}}{\phi^{*2}},
\end{eqnarray*}
with probability approaching one. Combining these inequalities with
(\ref{pf:2-2}), we have 
\[
\frac{1}{nb_{n}}|\mathbf{G}(\hat{\theta}-\theta^{\ast})|_{2}^{2}+\lambda_{n}|\hat{\theta}_{2}-\theta_{2}^{\ast}|_{1}\leq5\lambda_{n}^{2}\frac{s^{*}}{\phi^{*2}},
\]
which implies
\begin{equation}
|\hat{\theta}_{2}-\theta_{2}^{\ast}|_{1}\leq5\lambda_{n}\frac{s^{*}}{\phi^{*2}}.\label{pf:2-3}
\end{equation}

Turning to the finite dimensional component $\hat{\theta}_{1}$, note
that 
\begin{eqnarray*}
\hat{\theta}_{1}-\theta_{1}^{*} & = & \left(\frac{1}{nb_{n}}\mathbf{G}_{1}^{\prime}\mathbf{G}_{1}\right)^{-1}\left(\frac{1}{nb_{n}}\mathbf{G}_{1}^{\prime}e-\frac{1}{nb_{n}}\mathbf{G}_{1}^{\prime}\mathbf{G}_{2}(\hat{\theta}_{2}-\theta_{2}^{\ast})\right)\\
 & = & O_{p}((nb_{n})^{-1/2})+O_{p}(|\hat{\theta}_{2}-\theta_{2}^{\ast}|_{1}).
\end{eqnarray*}
Combining this with (\ref{pf:2-3}) yields the conclusion in (\ref{eq:l1b}).

\subsubsection{Proof of (ii) }

Note that we assume $h_{n}=b_{n}$. Let $\mathbf{G}_{\hat{S}}=(G_{\hat{S},1},\ldots,G_{\hat{S},n})^{\prime}$.
Since $\bar{\theta}_{\hat{S}}$ is the OLS estimator that minimizes
the sum of the squared residuals in the regression of $\mathbf{Y}$
on $\mathbf{G}_{\hat{S}}$ , it holds 
\[
|\mathbf{Y}-\mathbf{G}_{\hat{S}}\bar{\theta}_{\hat{S}}|_{2}^{2}\leq|\mathbf{Y}-\mathbf{G}_{\hat{S}}\hat{\theta}_{\hat{S}}|_{2}^{2}.
\]
Let $\hat{S}_{1}=\{i:0<|\hat{\theta}_{i}|<\zeta_{n}\}$ and $S_{n}=\hat{S}\cup\hat{S}_{1}$.
Note also that
\[
|\mathbf{Y}-\mathbf{G}_{S_{n}}\bar{\theta}_{S_{n}}|_{2}^{2}=|\mathbf{Y}-\mathbf{G}_{\hat{S}}\bar{\theta}_{\hat{S}}|_{2}^{2}\leq|\mathbf{Y}-\mathbf{G}_{\hat{S}}\hat{\theta}_{\hat{S}}|_{2}^{2}.
\]
Then, we get
\begin{align}
\frac{1}{nh_{n}}|\mathbf{G}_{S_{n}}(\bar{\theta}_{S_{n}}-\hat{\theta}_{S_{n}})|_{2}^{2} & \leq\frac{1}{nh_{n}}|\mathbf{G}_{\hat{S}_{1}}\hat{\theta}_{\hat{S}_{1}}|_{2}^{2}+\frac{2}{nh_{n}}|\hat{\mathbf{e}}^{\prime}\mathbf{G}_{\hat{S}_{1}}\hat{\theta}_{\hat{S}_{1}}|+\frac{2}{nh_{n}}|\hat{\mathbf{e}}^{\prime}\mathbf{G}_{S_{n}}(\bar{\theta}_{S_{n}}-\hat{\theta}_{S_{n}})|\nonumber \\
 & \leq\frac{1}{nh_{n}}|\mathbf{G}_{\hat{S}_{1}}\hat{\theta}_{\hat{S}_{1}}|_{2}^{2}+\lambda_{n}|\hat{\theta}_{\hat{S}_{1}}|_{1}+\lambda_{n}|\bar{\theta}_{S_{n}}-\hat{\theta}_{S_{n}}|_{1},\label{eq:311}
\end{align}
due to the Hölder inequality and the Karush\textendash Kuhn\textendash Tucker
condition for $\hat{\mathbf{e}}$. On the other hand, 
\[
\lambda_{\min}\left(\frac{1}{nh_{n}}\mathbf{G}_{S}^{\prime}\mathbf{G}_{S}\right)|\bar{\theta}_{S}-\hat{\theta}_{S}|_{2}^{2}\leq\frac{1}{nh_{n}}|\mathbf{G}_{S}(\bar{\theta}_{S}-\hat{\theta}_{S})|_{2}^{2}.
\]
Also note that $|a|_{1}\leq\sqrt{s}|a|_{2}$ for an $s$-dimensional
vector $a$. Then, we consider the three cases. First, let the last
term in (\ref{eq:311}) be the biggest among the three terms. Then,
the bound becomes $\lambda_{\min}\left(\frac{1}{nh_{n}}\mathbf{G}_{S_{n}}^{\prime}\mathbf{G}_{S_{n}}\right)^{-1}|S_{n}|\lambda_{n}.$
If the second is the biggest, then $|\hat{S}_{1}|\zeta_{n}\geq|\hat{\theta}_{\hat{S}_{1}}|_{1}\geq|\bar{\theta}_{S_{n}}-\hat{\theta}_{S_{n}}|_{1}$.
Finally, if the first term is the biggest, then $\lambda_{n}|\bar{\theta}_{S_{n}}-\hat{\theta}_{S_{n}}|_{1}\leq\frac{1}{nh_{n}}|\mathbf{G}_{\hat{S}_{1}}\hat{\theta}_{\hat{S}_{1}}|_{2}^{2}\leq\lambda_{\max}\left(\frac{1}{nh_{n}}\mathbf{G}_{\hat{S}_{1}}^{\prime}\mathbf{G}_{\hat{S}_{1}}\right)|\hat{S}_{1}|\zeta_{n}^{2}$.
Therefore, we get the desired bound under $h_{n}=b_{n}$ and Assumption
\ref{asm:point-est}.

\subsection{Proof of Theorem \ref{thm:t} \label{sub:pf2}}

First, we show the asymptotic expansion in (\ref{eq:lin}). Let $\bar{\gamma}$
be a coefficient vector of $Z_{i}$, where the elements correspond
to $Z_{\hat{S},i}$ is $\bar{\gamma}_{\hat{S}}$ in (\ref{eq:post-lasso}),
and other elements are zero. Observe that
\begin{eqnarray*}
\bar{\tau} & = & e_{2}^{\prime}\left(\sum_{i=1}^{n}K\left(\frac{X_{i}}{h_{n}}\right)G_{1i}G_{1i}^{\prime}\right)^{-1}\sum_{i=1}^{n}K\left(\frac{X_{i}}{h_{n}}\right)G_{1i}(Y_{i}-Z_{i}^{\prime}\bar{\gamma})\\
 & = & \hat{\tau}_{\xi}-e_{2}^{\prime}\left(\sum_{i=1}^{n}K\left(\frac{X_{i}}{h_{n}}\right)G_{1i}G_{1i}^{\prime}\right)^{-1}\sum_{i=1}^{n}K\left(\frac{X_{i}}{h_{n}}\right)G_{1i}Z_{i}^{\prime}(\bar{\gamma}-\gamma_{Y}),
\end{eqnarray*}
where $\hat{\tau}_{\xi}=e_{2}^{\prime}\left(\sum_{i=1}^{n}K(X_{i}/h_{n})G_{1i}G_{1i}^{\prime}\right)^{-1}\sum_{i=1}^{n}K(X_{i}/h_{n})G_{1i}\xi_{i}$.
Thus it is sufficient for (\ref{eq:lin}) to show that the second
term is of order $o_{p}((nh_{n})^{-1/2})$.

Note that
\begin{eqnarray*}
 &  & \left\Vert e_{2}^{\prime}\left(\frac{1}{nh_{n}}\sum_{i=1}^{n}K\left(\frac{X_{i}}{h_{n}}\right)G_{1i}G_{1i}^{\prime}\right)^{-1}\frac{1}{\sqrt{nh_{n}}}\sum_{i=1}^{n}K\left(\frac{X_{i}}{h_{n}}\right)G_{1i}Z_{i}^{\prime}\right\Vert _{\infty}\\
 & \le & \lambda_{\min}\left(\frac{1}{nh_{n}}\sum_{i=1}^{n}K\left(\frac{X_{i}}{h_{n}}\right)G_{1i}G_{1i}^{\prime}\right)^{-1}\left\Vert \frac{1}{\sqrt{nh_{n}}}\sum_{i=1}^{n}\left\{ K\left(\frac{X_{i}}{h_{n}}\right)G_{1i}Z_{i}^{\prime}-\mathbb{E}\left[K\left(\frac{X_{i}}{h_{n}}\right)G_{1i}Z_{i}^{\prime}\right]\right\} \right\Vert _{\infty}\\
 &  & +O_{p}(\sqrt{nh_{n}}h_{n}^{2})\\
 & = & O_{p}(\sqrt{\log p})+O_{p}(\sqrt{nh_{n}}h_{n}^{2}),
\end{eqnarray*}
where the inequality follows from the fact that $\left(\sum_{i=1}^{n}K(X_{i}/h_{n})G_{1i}G_{1i}^{\prime}\right)^{-1}\sum_{i=1}^{n}K(X_{i}/h_{n})G_{1i}Z_{i}^{\prime}$
is a vector of the local linear RDD estimator for the outcome $Z_{i}$
and its bias is of order $O(h_{n}^{2})$ from Lemma SA-2 of CCFT,
and the equality follows from the local Bernstein's inequality in
Lemma \ref{lem:Bernstein}. 

Turing to bound $\bar{\gamma}-\gamma_{Y}$, recall that $\theta_{S^{*}}^{*}=\arg\min_{\theta_{S^{*}}}\mathbb{E}[K(X_{i}/b_{n})(Y_{i}-G_{S^{*}i}^{\prime}\theta_{S^{*}})^{2}/b_{n}]$
while $\gamma_{Y}=\arg\min_{\gamma}\mathbb{E}[(\tilde{Y}-Z_{S^{*}}^{\prime}\gamma)^{2}|X=0]$,
where $\tilde{Y}=Y(1)-Y(0)-\mathbb{E}[Y(1)-Y(0)|X=0]$ and $Z_{S^{*}}=Z_{S^{*}}(1)-Z_{S^{*}}(0)$.
Due to van der Vaart and Wellner (1996, Theorem 3.4.1), we can bound
the contrast $|\gamma_{Y}-\gamma^{*}|$ by the difference in the criterions,
which is of order $O(s^{*}b_{n}^{2})$ by the standard bias calculation.
For the same reasoning, we obtain $|\bar{\gamma}(b_{n})-\bar{\gamma}(h_{n})|=O_{p}(s^{*}(h_{n}^{2}+b_{n}^{2}))$,
where we highlight by $\bar{\gamma}(b_{n})$ the dependence on the
bandwidth used to compute $\bar{\gamma}$.

Since $|\bar{\gamma}(b_{n})-\gamma^{*}|_{1}=O_{p}(\lambda_{n}s^{*})$
by Theorem \ref{thm:point-est} and thus $|\gamma_{Y}-\bar{\gamma}(h_{n})|_{1}=O_{p}(s^{*}(h_{n}^{2}+b_{n}^{2})+\lambda_{n}s^{*})$
due to the preceding derivation, Hölder's inequality and the assumption
$(\sqrt{\log p}+\sqrt{nh_{n}}h_{n}^{2})(s^{*}(h_{n}^{2}+b_{n}^{2})+\lambda_{n}s^{*})\to0$
guarantee (\ref{eq:lin}).

Second, we note that $\hat{\tau}_{\xi}$ is the conventional local
linear RDD estimator without covariates for the outcome variable $\xi_{i}$.
Thus, the proof of CCFT's Theorem 1 is directly applicable and the
MSE expansion in (\ref{eq:MSE}) follows.

Finally, we show (\ref{eq:tau-t}). By (\ref{eq:lin}), we have 
\[
T_{\tau}=\sqrt{\frac{\mathcal{V}}{\bar{\mathcal{V}}}}\sqrt{\frac{nh_{n}}{\mathcal{V}}}(\hat{\tau}_{\xi}-h_{n}^{2}\mathcal{B}-\tau)-\sqrt{\frac{\mathcal{V}}{\bar{\mathcal{V}}}}\sqrt{\frac{nh_{n}}{\mathcal{V}}}h_{n}^{2}(\bar{\mathcal{B}}-\mathcal{B}).
\]
Since $\hat{\tau}_{\xi}$ is the conventional local linear RDD estimator
without covariates for the outcome variable $\xi_{i}$, Lemma SA-10
of CCFT yields $\sqrt{\frac{nh_{n}}{\mathcal{V}}}(\hat{\tau}_{\xi}-h_{n}^{2}\mathcal{B}-\tau)\overset{d}{\to}N(0,1)$.
Therefore, the assumptions $\sqrt{\frac{nh_{n}^{5}}{\mathcal{V}}}(\bar{\mathcal{B}}-\mathcal{B})\overset{p}{\to}0$
and $\frac{\bar{\mathcal{V}}}{\mathcal{V}}\overset{p}{\to}1$ imply
the conclusion in (\ref{eq:tau-t}).

\subsection{Proof of Proposition \ref{prop:V}}

Observe that $\bar{\mathcal{V}}^{NN}$ can be written as
\[
\bar{\mathcal{V}}^{NN}=\bar{r}_{-}^{\prime}\bar{\Psi}_{-}^{NN}\bar{r}_{-}+\bar{r}_{+}^{\prime}\bar{\Psi}_{+}^{NN}\bar{r}_{+},
\]
where $\bar{r}_{-}=\bar{q}\otimes\Gamma_{-}^{-1}e_{1}$, $\bar{r}_{+}=\bar{q}\otimes\Gamma_{+}^{-1}e_{1}$,
and 
\begin{eqnarray*}
\bar{\Psi}_{-}^{NN} & = & \left[\begin{array}{cccc}
\bar{\Psi}_{YY-}^{NN} & \bar{\Psi}_{YZ_{1}-}^{NN} & \cdots & \bar{\Psi}_{YZ_{|\hat{S}|}-}^{NN}\\
\bar{\Psi}_{Z_{1}Y-}^{NN} & \bar{\Psi}_{Z_{1}Z_{1}-}^{NN}\\
\vdots &  & \ddots\\
\bar{\Psi}_{Z_{|\hat{S}|}Y-}^{NN} &  &  & \bar{\Psi}_{Z_{|\hat{S}|}Z_{|\hat{S}|}-}^{NN}
\end{array}\right],\quad\bar{\Psi}_{+}^{NN}=\left[\begin{array}{cccc}
\bar{\Psi}_{YY+}^{NN} & \bar{\Psi}_{YZ_{1}+}^{NN} & \cdots & \bar{\Psi}_{YZ_{|\hat{S}|}+}^{NN}\\
\bar{\Psi}_{Z_{1}Y+}^{NN} & \bar{\Psi}_{Z_{1}Z_{1}+}^{NN}\\
\vdots &  & \ddots\\
\bar{\Psi}_{Z_{|\hat{S}|}Y+}^{NN} &  &  & \bar{\Psi}_{Z_{|\hat{S}|}Z_{|\hat{S}|}+}^{NN}
\end{array}\right],\\
\bar{\Psi}_{VW-}^{NN} & = & \frac{1}{nh_{n}}\sum_{i=1}^{n}\mathbb{I}\{X_{i}<0\}K(X_{i}/h_{n})^{2}\left[\begin{array}{cc}
1 & X_{i}/h_{n}\\
X_{i}/h_{n} & (X_{i}/h_{n})^{2}
\end{array}\right]\bar{\varepsilon}_{V-,i}\bar{\varepsilon}_{W-,i},\\
\bar{\Psi}_{VW+}^{NN} & = & \frac{1}{nh_{n}}\sum_{i=1}^{n}\mathbb{I}\{X_{i}\ge0\}K(X_{i}/h_{n})^{2}\left[\begin{array}{cc}
1 & X_{i}/h_{n}\\
X_{i}/h_{n} & (X_{i}/h_{n})^{2}
\end{array}\right]\bar{\varepsilon}_{V+,i}\bar{\varepsilon}_{W+,i},
\end{eqnarray*}
for $V,W\in\{Y,Z_{1},\ldots,Z_{p}\}$. Similarly, the asymptotic variance
$\mathcal{V}$ can be written as
\[
\mathcal{V}=r_{-}^{\prime}\Psi_{-}r_{-}+r_{+}^{\prime}\Psi_{+}r_{+},
\]
where $r_{-}=q\otimes\Gamma_{-}^{-1}e_{1}$, $r_{+}=q\otimes\Gamma_{+}^{-1}e_{1}$,
and 
\begin{eqnarray*}
\Psi_{-} & = & \left[\begin{array}{cccc}
\Psi_{YY-} & \Psi_{YZ_{1}-} & \cdots & \Psi_{YZ_{s^{*}}-}\\
\Psi_{Z_{1}Y-} & \Psi_{Z_{1}Z_{1}-}\\
\vdots &  & \ddots\\
\Psi_{Z_{s^{*}}Y-} &  &  & \Psi_{Z_{s^{*}}Z_{s^{*}}-}
\end{array}\right],\quad\Psi_{+}=\left[\begin{array}{cccc}
\Psi_{YY+} & \Psi_{YZ_{1}+} & \cdots & \Psi_{YZ_{s^{*}}+}\\
\Psi_{Z_{1}Y+} & \Psi_{Z_{1}Z_{1}+}\\
\vdots &  & \ddots\\
\Psi_{Z_{s^{*}}Y+} &  &  & \Psi_{Z_{s^{*}}Z_{s^{*}}+}
\end{array}\right],\\
\Psi_{VW-} & = & \frac{1}{nh_{n}}\sum_{i=1}^{n}\mathbb{I}\{X_{i}<0\}K(X_{i}/h_{n})^{2}\left[\begin{array}{cc}
1 & X_{i}/h_{n}\\
X_{i}/h_{n} & (X_{i}/h_{n})^{2}
\end{array}\right]\mathrm{Cov}(V_{i}(0),W_{i}(0)|X_{i}),\\
\Psi_{VW+} & = & \frac{1}{nh_{n}}\sum_{i=1}^{n}\mathbb{I}\{X_{i}\ge0\}K(X_{i}/h_{n})^{2}\left[\begin{array}{cc}
1 & X_{i}/h_{n}\\
X_{i}/h_{n} & (X_{i}/h_{n})^{2}
\end{array}\right]\mathrm{Cov}(V_{i}(1),W_{i}(1)|X_{i}),
\end{eqnarray*}
for $V,W\in\{Y,Z_{1},\ldots,Z_{p}\}$.

Since $\mathbb{P}\{\hat{S}\subset S^{*}\}\to1$ due to the deviation
bound in Theorem 1, the following statements are all conditional on
the event $\{\hat{S}\subset S^{*}\}$. Without loss of generality,
suppose $\hat{S}=\{1,\ldots,|\hat{S}|\}$ and $S^{*}=\{1,\ldots,|\hat{S}|,|\hat{S}|+1,\ldots,s^{*}\}$.
Let
\begin{equation}
\varpi_{n}=\min\{\lambda_{\min}(\Psi_{-}),\lambda_{\min}(\Psi_{+})\}.\label{pf:omega}
\end{equation}
 We decompose
\begin{eqnarray*}
\frac{\bar{\mathcal{V}}^{NN}}{\mathcal{V}}-1 & = & \frac{1}{\mathcal{V}}\left[\begin{array}{c}
\bar{r}_{-}\\
0
\end{array}\right]^{\prime}\left(\left[\begin{array}{cc}
\bar{\Psi}_{-}^{NN} & 0\\
0 & 0
\end{array}\right]-\Psi_{-}\right)\left[\begin{array}{c}
\bar{r}_{-}\\
0
\end{array}\right]+\frac{1}{\mathcal{V}}\left(\left[\begin{array}{c}
\bar{r}_{-}\\
0
\end{array}\right]+r_{-}\right)^{\prime}\Psi_{-}\left(\left[\begin{array}{c}
\bar{r}_{-}\\
0
\end{array}\right]+r_{-}\right)\\
 &  & +\frac{1}{\mathcal{V}}\left[\begin{array}{c}
\bar{r}_{+}\\
0
\end{array}\right]^{\prime}\left(\left[\begin{array}{cc}
\bar{\Psi}_{+}^{NN} & 0\\
0 & 0
\end{array}\right]-\Psi_{+}\right)\left[\begin{array}{c}
\bar{r}_{+}\\
0
\end{array}\right]+\frac{1}{\mathcal{V}}\left(\left[\begin{array}{c}
\bar{r}_{+}\\
0
\end{array}\right]+r_{+}\right)^{\prime}\Psi_{+}\left(\left[\begin{array}{c}
\bar{r}_{+}\\
0
\end{array}\right]+r_{+}\right).\\
 & =: & T_{1-}+T_{2-}+T_{1+}+T_{2+}.
\end{eqnarray*}
For $T_{2-}$, we have
\begin{eqnarray*}
|T_{2-}| & \le & \varpi_{n}^{-1}\left|\left[\begin{array}{c}
\bar{r}_{-}\\
0
\end{array}\right]+r_{-}\right|_{1}\left|\Psi_{-}\left(\left[\begin{array}{c}
\bar{r}_{-}\\
0
\end{array}\right]+r_{-}\right)\right|_{\infty}\\
 & = & O_{p}\left(\varpi_{n}^{-1}\{(m_{n}|S_{n}|\lambda_{n})\vee(|\hat{S}_{1}|\zeta_{n})\vee(m_{1n}|\hat{S}_{1}|\zeta_{n}^{2}/\lambda_{n})\}\right),
\end{eqnarray*}
where the inequality follows from $\mathcal{V}\ge\varpi_{n}$ and
Hölder's inequality, and the equality follows from Theorem \ref{thm:point-est}
(ii). A similar argument yields that $T_{2+}$ is of the same stochastic
order. 

For $T_{1-}$, letting $\Psi_{-}^{11}$ be the first $|\hat{S}|\times|\hat{S}|$
block component of $\Psi_{-}$, we have
\begin{eqnarray*}
|T_{1-}| & \le & \varpi_{n}^{-1}\lambda_{\max}(\bar{\Psi}_{-}^{NN}-\Psi_{-}^{11})\bar{r}_{-}^{\prime}\bar{r}_{-}\le\max_{1\le j_{1},j_{2}\le2\hat{S}}|\bar{\Psi}_{-,l_{1},l_{2}}^{NN}-\Psi_{-,l_{1},l_{2}}^{11}|O_{p}(\varpi_{n}^{-1}s^{*}).
\end{eqnarray*}
So it remains to obtain the stochastic order of $\max_{1\le l_{1},l_{2}\le2\hat{S}}|\bar{\Psi}_{-,l_{1},l_{2}}^{NN}-\Psi_{-,l_{1},l_{2}}^{11}|$.
From p. 36 of the supplement of CCT, we can decompose
\[
\bar{\Psi}_{-,VW}^{NN}-\Psi_{-,VW}^{11}=\eta_{1}^{VW}+\eta_{2}^{VW}+\eta_{3}^{VW},
\]
for $V,W\in\{Y,Z_{1},\ldots,Z_{p}\}$, where $\varepsilon_{V,i}=V_{i}-\mathbb{E}[V_{i}|X_{i}]$,
$\varepsilon_{W,i}=W_{i}-\mathbb{E}[W_{i}|X_{i}]$, $\mu_{V-}(x)=\mathbb{E}[V_{i}(0)|X_{i}=x]$,
$\mu_{W-}(x)=\mathbb{E}[W_{i}(0)|X_{i}=x]$, and 
\begin{eqnarray*}
\eta_{1}^{VW} & = & \frac{1}{J(J+1)}\sum_{j=1}^{J}\eta_{1,j}^{VW},\\
\eta_{1,j}^{VW} & = & \frac{1}{nh_{n}}\sum_{i=1}^{n}\mathbb{I}\{X_{i}<0\}K(X_{i}/h_{n})^{2}\left[\begin{array}{cc}
1 & X_{i}/h_{n}\\
X_{i}/h_{n} & (X_{i}/h_{n})^{2}
\end{array}\right](\varepsilon_{V,\ell_{-,j}(i)}\varepsilon_{W,\ell_{-,j}(i)}-\varepsilon_{V,i}\varepsilon_{W,i}),\\
\eta_{2}^{VW} & = & \frac{2}{J(J+1)}\sum_{1\le j,k\le J}\eta_{2,j,k}^{VW},\\
\eta_{2,j,k}^{VW} & = & \frac{1}{n}\sum_{i=1}^{n}\mathbb{I}\{X_{i}<0\}K(X_{i}/h_{n})^{2}\left[\begin{array}{cc}
1 & X_{i}/h_{n}\\
X_{i}/h_{n} & (X_{i}/h_{n})^{2}
\end{array}\right]\varepsilon_{V,\ell_{-,j}(i)}\varepsilon_{W,\ell_{-,k}(i)},\\
\eta_{3}^{VW} & = & \frac{1}{n}\sum_{i=1}^{n}\mathbb{I}\{X_{i}<0\}K(X_{i}/h_{n})^{2}\left[\begin{array}{cc}
1 & X_{i}/h_{n}\\
X_{i}/h_{n} & (X_{i}/h_{n})^{2}
\end{array}\right]\\
 &  & \qquad\times\{(\hat{\sigma}_{3,i}^{VW})^{2}-(\hat{\sigma}_{4,i}^{VW})^{2}-(\hat{\sigma}_{5,i}^{VW})^{2}+(\hat{\sigma}_{6,i}^{VW})^{2}+(\hat{\sigma}_{7,i}^{VW})^{2}-(\hat{\sigma}_{8,i}^{VW})^{2}-(\hat{\sigma}_{9,i}^{VW})^{2}\},\\
(\hat{\sigma}_{3,i}^{VW})^{2} & = & \frac{1}{J(J+1)}\left(\sum_{j=1}^{J}\{\mu_{V-}(X_{i})-\mu_{V-}(X_{\ell_{-,j}(i)})\}\right)\left(\sum_{j=1}^{J}\{\mu_{W-}(X_{i})-\mu_{W-}(X_{\ell_{-,j}(i)})\}\right),\\
(\hat{\sigma}_{4,i}^{VW})^{2} & = & \varepsilon_{V,i}\frac{1}{J+1}\sum_{j=1}^{J}\varepsilon_{W,\ell_{-,j}(i)},\qquad(\hat{\sigma}_{5,i}^{VW})^{2}=\varepsilon_{W,i}\frac{1}{J+1}\sum_{j=1}^{J}\varepsilon_{V,\ell_{-,j}(i)},\\
(\hat{\sigma}_{6,i}^{VW})^{2} & = & \varepsilon_{V,i}\frac{1}{J+1}\sum_{j=1}^{J}\{\mu_{W-}(X_{i})-\mu_{W-}(X_{\ell_{-,j}(i)})\},\\
(\hat{\sigma}_{7,i}^{VW})^{2} & = & \varepsilon_{W,i}\frac{1}{J+1}\sum_{j=1}^{J}\{\mu_{V-}(X_{i})-\mu_{V-}(X_{\ell_{-,j}(i)})\},\\
(\hat{\sigma}_{8,i}^{VW})^{2} & = & \frac{1}{J+1}\sum_{j=1}^{J}\varepsilon_{V,\ell_{-,j}(i)}\sum_{k=1}^{J}\{\mu_{W-}(X_{i})-\mu_{W-}(X_{\ell_{-,j}(i)})\},\\
(\hat{\sigma}_{9,i}^{VW})^{2} & = & \frac{1}{J+1}\sum_{j=1}^{J}\varepsilon_{W,\ell_{-,j}(i)}\sum_{k=1}^{J}\{\mu_{V-}(X_{i})-\mu_{V-}(X_{\ell_{-,j}(i)})\}.
\end{eqnarray*}
Since the argument is similar as the one in Section S.2.4 of CCT's
supplement, we only present the result for $\eta_{1}^{VW}$. Since
$J$ is fixed, take any $j$. We want to obtain the stochastic order
of
\[
\max_{V,W\in\{Y,Z_{1},\ldots,Z_{|\hat{S}|}\}}\left|\frac{1}{nh_{n}}\sum_{i=1}^{n}\mathbb{I}\{X_{i}<0\}K(X_{i}/h_{n})^{2}(\varepsilon_{V,\ell_{-,j}(i)}\varepsilon_{W,\ell_{-,j}(i)}-\varepsilon_{V,i}\varepsilon_{W,i})\right|.
\]
Note that
\begin{eqnarray*}
 &  & \max_{V,W\in\{Y,Z_{1},\ldots,Z_{|\hat{S}|}\}}\left|\frac{1}{nh_{n}}\sum_{i=1}^{n}\mathbb{I}\{X_{i}<0\}K(X_{i}/h_{n})^{2}(\varepsilon_{V,\ell_{-,j}(i)}\varepsilon_{W,\ell_{-,j}(i)}-\varepsilon_{V,i}\varepsilon_{W,i})\right|\\
 & \le & \max_{V,W\in\{Y,Z_{1},\ldots,Z_{|\hat{S}|}\}}\left|\frac{1}{nh_{n}}\sum_{i=1}^{n}\mathbb{I}\{X_{i}<0\}K(X_{i}/h_{n})^{2}\left\{ \begin{array}{c}
(\varepsilon_{V,\ell_{-,j}(i)}\varepsilon_{W,\ell_{-,j}(i)}-\varepsilon_{V,i}\varepsilon_{W,i})\\
-(\sigma_{VW}^{2}(X_{\ell_{-,j}(i)})-\sigma_{VW}^{2}(X_{i}))
\end{array}\right\} \right|\\
 &  & +\max_{V,W\in\{Y,Z_{1},\ldots,Z_{|\hat{S}|}\}}\left|\frac{1}{nh_{n}}\sum_{i=1}^{n}\mathbb{I}\{X_{i}<0\}K(X_{i}/h_{n})^{2}(\sigma_{VW}^{2}(X_{\ell_{-,j}(i)})-\sigma_{VW}^{2}(X_{i}))\right|\\
 & =: & T_{11}+T_{12}.
\end{eqnarray*}
For $T_{11}$, we can apply the localized Bernstein inequality in
Lemma \ref{lem:Bernstein}, which implies
\[
T_{11}=O_{p}\left(\sqrt{\frac{\log s^{*}}{nh_{n}}}\right).
\]
For $T_{12}$, note that
\begin{eqnarray*}
T_{12} & = & \max_{V,W\in\{Y,Z_{1},\ldots,Z_{|\hat{S}|}\}}\left|\frac{1}{nh_{n}}\sum_{i=1}^{n}\mathbb{I}\{X_{i}<0\}K(X_{i}/h_{n})^{2}(\sigma_{VW}^{2}(X_{\ell_{j}^{-}(i)})-\sigma_{VW}^{2}(X_{i}))\right|\\
 & \le & \max_{1\le i\le n}|X_{\ell_{j}^{-}(i)}-X_{i}|\max_{V,W\in\{Y,Z_{1},\ldots,Z_{|\hat{S}|}\}}|C_{VW}|\left|\frac{1}{nh_{n}}\sum_{i=1}^{n}\mathbb{I}\{X_{i}<0\}K(X_{i}/h_{n})^{2}\right|\\
 & = & O_{p}\left(\max_{1\le i\le n}|X_{\ell_{j}^{-}(i)}-X_{i}|\right)=O_{p}(n^{-1}),
\end{eqnarray*}
where $C_{VW}$ is the Lipschitz constant for $\sigma_{VW}^{2}$ (and
we assume $\max_{V,W\in\{Y,Z_{1},\ldots,Z_{|\hat{S}|}\}}|C_{VW}|$
is bounded), and the second equality follows from Abadie and Imbens
(2006, Theorem 1). Combining these results, we have
\[
T_{1-}=O_{p}\left(\frac{s^{*}\sqrt{\log s^{*}}}{\varpi_{n}nh_{n}}\right).
\]

\newpage{}

\end{document}